\documentclass[a4paper,fleqn,usenatbib]{mnras}

\usepackage{newtxtext,newtxmath}

\usepackage[T1]{fontenc}
\usepackage{ae,aecompl}

\usepackage{graphicx}	
\usepackage{amsmath}	
\usepackage{amssymb}	
\usepackage{siunitx}
\usepackage{upgreek}
\usepackage{epstopdf}

\usepackage{array}
\newcolumntype{H}{>{\setbox0=\hbox\bgroup}c<{\egroup}@{}}

\newcommand{\spie}{SPIE}
\newcommand{\pnas}{PNAS}

\title[Constraints on planets in two-belt systems]{Constraining the presence of giant planets in two-belt debris disk systems with VLT/SPHERE direct imaging and dynamical arguments}

\author[E. C. Matthews et al.]{
Elisabeth Matthews$^{1}$\thanks{E-mail: ematthews@astro.ex.ac.uk},
Sasha Hinkley$^{1}$, 
Arthur Vigan$^{2}$, 
Grant Kennedy$^{3}$,
\newauthor
Ben Sutlieff$^{1}$,
Dawn Wickenden$^{1}$,
Sam Treves$^{1}$,
Trevor David$^{4}$,
Tiffany Meshkat$^{5,4}$, 
\newauthor 
Dimitri Mawet$^{4,6}$,
Farisa Morales$^{4}$,
Andrew Shannon$^{7,8}$,
Karl Stapelfeldt$^{4}$
\\
$^{1}$University of Exeter, Physics Department, Stocker Road, Exeter, EX4 4QL, UK \\
$^{2}$Aix Marseille Universit\'e, CNRS, LAM (Laboratoire d'Astrophysique de Marseille) UMR 7326, 13388 Marseille, France \\
$^{3}$Department of Physics, University of Warwick, Gibbet Hill Road, Coventry CV4 7AL, UK \\
$^{4}$Jet Propulsion Laboratory, California Institute of Technology, 4800 Oak Grove Drive, Pasadena, CA 91109, USA \\
$^{5}$IPAC, Caltech, M/C 100-22, 1200 East California Boulevard, Pasadena, CA 91125, USA \\
$^{6}$Department of Astronomy, California Institute of Technology, 1200 E. California Boulevard, MC 249-17, Pasadena, CA 91125, USA \\
$^{7}$Department of Astronomy and Astrophysics, 525 Davey Laboratory, Pennsylvania State University, University Park, PA 16802, USA \\
$^{8}$Center for Exoplanets and Habitable Worlds, The Pennsylvania State University, State College, PA 16802, USA
}

\date{\vspace{-1cm}}

\pubyear{2018}

\begin{document}
\label{firstpage}
\pagerange{\pageref{firstpage}--\pageref{lastpage}}
\maketitle

\begin{abstract}

Giant, wide-separation planets often lie in the gap between multiple, distinct rings of circumstellar debris: this is the case for the HR\,8799 and HD\,95086 systems, and even the solar system where the Asteroid and Kuiper belts enclose the four gas and ice giants. In the case that a debris disk, inferred from an infrared excess in the SED, is best modelled as two distinct temperatures, we infer the presence of two spatially separated rings of debris. Giant planets may well exist between these two belts of debris, and indeed could be responsible for the formation of the gap between these belts. We observe 24 such two-belt systems using the VLT/SPHERE high contrast imager, and interpret our results under the assumption that the gap is indeed formed by one or more giant planets. A theoretical minimum mass for each planet can then be calculated, based on the predicted dynamical timescales to clear debris. The typical dynamical lower limit is $\sim$0.2$M_J$ in this work, and in some cases exceeds 1$M_J$. Direct imaging data, meanwhile, is typically sensitive to planets down to $\sim$3.6$M_J$ at 1'', and 1.7$M_J$ in the best case. Together, these two limits tightly constrain the possible planetary systems present around each target, many of which will be detectable with the next generation of high-contrast imagers.

\end{abstract}


\vspace{-1cm}

\section{Introduction}

Directly imaged planets are rare. This has been demonstrated by numerous surveys over the last decade: NaCo \citep{chauvin2015}, the Lyot project \citep{leconte2010}, GDPS \citep{lafreniere2007}, IDPS \citep{galicher2016}, SEEDS \citep{brandt2014}, NICI \citep{biller2013,nielsen2013,wahhaj2013} and others. In a meta-analysis of several deep imaging surveys, \citet{bowler2016} found an overall occurrence rate of $0.6^{+0.7}_{-0.5}$\% for companions in the range 5-13$M_J$ and 30-300~AU. \citet{galicher2016}, meanwhile, use a slightly wider parameter space of 0.5-14$M_J$ and 20-300~AU, and find an occurrence rate of $1.05^{+2.80}_{-1.70}$\%. Directly imaged planets are rare in the modest region of parameter space that can be probed, i.e. the most massive planets at the widest of separations.

The latest generation of direct imagers \citep[notably SPHERE and GPI, see][respectively]{beuzit2008,macintosh2014} are sensitive to lower masses of wide-separation planets that were previously inaccessible to direct imaging. These instruments are proving to have excellent high contrast abilities, and the GPIES and SPHERE/SHINE surveys are initially consistent with the low occurrence rates for wide-separation planets within the region where direct imaging is sensitive, i.e. massive planets in the Jupiter-mass regime, separated by tens of AU from their host stars. Only a few planets have been identified with these instruments so far: the GPIES team detected a planet around the $\beta$-Pictoris member 51 Eridani \citep{macintosh2015,derosa2015}, while a planet around the Sco-Cen star HIP~65426 has been detected by the SPHERE/SHINE team \citep{chauvin2017}. An exoplanet PDS~70b was very recently identified with SPHERE, in a gap within the transitional disk of this object \citep{keppler2018}.

In contrast to this relatively small number of planet detections, great success has been had with both SPHERE and GPI in detecting and characterizing debris dust systems in scattered light \citep[e.g.][]{currie2015b,kasper2015,draper2016,wahhaj2016,feldt2017,bonnefoy2017,matthews2017}. These systems are of particular interest since the presence of dust in a system may correlate with the presence of planets. Dust is transient, being blown out of systems by stellar winds or falling onto the stellar surface via the Poynting-Robertson effect. Therefore, if dust is observed to be present it must be constantly regenerated via planetesimal collisions. Planetesimals are the building blocks of planets, and so their presence is a useful indicator that planets may also have been able to form in a certain system. Even further, the presence of one or more giant planets in a system may perturb the orbits of these planetesimals, further increasing the rate of dust production \citep{mustillwyatt2009}. Those systems that host massive, wide-orbit planets might therefore also show evidence for particularly high quantities of dust.

Many of the known directly imaged planets reside in highly dusty systems. For example, the massive debris disk around $\beta$-Pictoris was first imaged by \citet{smith1984}, and a massive planet was subsequently detected by \citet{lagrange2009}. It is worth noting however that this correlation does not itself imply an underlying link between debris disks and planetary systems, since many directly imaged planets have been discovered in surveys deliberately targeting a biased selection of highly dusty disks. Nonetheless, \citet{meshkat2017} found that there is a statistically significant excess (at the 88\% confidence level) of planets around highly dusty stars, compared to the occurence rates in a control sample, for early type stars.

These dusty systems also allow the study of the dynamical interactions of dust and planets. A sharp disk edge or a gap between two belts of debris dust can be formed by the gravitational influence of a giant planet. This has been observed in the HR~8799 system \citep{marois2008,marois2010} which hosts four known planets, with radii between 14~AU and 68~AU, and two distinct debris belts at $\sim$9~AU and beyond $\sim$95~AU \citep{reidemeister2009,su2009,matthews_b2014}. HD~95086 shows similar system architecture, with two distinct debris belts \citep{su2015} and one known planet \citep{rameau2013} lying between them. \citet{su2015} present possible architectures for this system with up to four planets clearing the gap between these debris belts, the inner three being below current detection limits. Even our own solar system is in this configuration, with the Asteroid and Kuiper belts enclosing four large, wide-separation gas and ice giants.

Systems in this two-belt configuration can be detected by observations of an infrared excess: if this infrared excess is best modelled as two distinct temperatures, as is the case for both HR~8799 and HD~95086, we infer that there are two temperatures of dust and therefore probably rings of dust at two radii (see e.g. \citealt{kennedy2014}). These two-belt systems are unique in that there is spatial information suggesting \textit{where} in the system planets are likely to be found. By assuming that the debris gap is formed by the gravitational clearing of one or more giant planets, we conclude that the planets in these systems should lie between the inner and outer debris belt radii, as inferred from infrared SED fitting. Under the assumption that planets are equal mass and typically separated by $\sim$20 mutual Hill radii, it is even possible to deduce the predicted location of each individual planet in a multi-planet system, based on the number of planets we expect. The mutual Hill radius is defined as 
\begin{equation}
    R_H = \frac{a_1+a_2}{2} \times \left(\frac{m_1+m_2}{3M_\star}\right)^{\frac{1}{3}}
    \label{eq:mutualhillradius}
\end{equation}
for planets with masses $m_1$ and $m_2$, and semi-major axes $a_1$ and $a_2$. For transiting planets observed with \textit{Kepler}, \citet{fangmargot2013} found a typical planet-planet separation of 21.7$\pm$9.5$R_H$. While there is no guarantee that massive, wide-separation planets will behave as close-in planets do, we note that a significantly closer spacing is likely unstable. The HR~8799 planets are separated by as little as 3-4 mutual Hill radii, but the system is only stable due to the special dynamical configuration of the planets with several mean motion resonances \citep{fabryckymurrayclay2010,gozdziewskimigaszewski2014}.

As well as using disk structure to predict the locations of planets, it is possible to use dynamical arguments to constrain exoplanet masses. By assuming several equal mass planets spread across a debris gap, \citet{shannon2016} found that the clearing time scales with the planet mass and the width of the debris gap. For a system with widely spaced debris belts and giant planets, this timescale is of order millions of years, and as such is similar to the lifetime of the system. By imposing that the clearing time be less than the stellar age, it is possible to calculate the minimum mass of each planet in the system that would facilitate clearing of the observed debris gap. This constraint can be combined with upper mass limits based on direct imaging analysis, so as to place tight limits on the possible planetary configurations in these multi-belt systems.

In this work, we survey 24 systems with previously published evidence for debris disks segregated into two distinct belts. We search for evidence for the planets that might be responsible for sculpting these debris disks, and test how tightly the undetected planetary systems can be constrained. Section \ref{targets} describes our target selection, and our observations and data reduction are discussed in sections \ref{obs} and \ref{datared} respectively. The contrast limits and candidate companion identification are given in section \ref{results}, and we discuss our results in section \ref{analysis}.

\section{Target Selection} \label{targets}

For this survey the aim was to study systems hosting the best characterized multi-belt debris disks, as determined by fitting of the infrared excess emission. To do this, targets were selected that are presented in \citet{chen2014} as hosting two-temperature debris disks. However, fitting the infrared excess is inherently complex, and there are often disagreements in the literature about the nature of a certain target. Many of our targets appear in the literature in \citet{morales2011}, \citet{ballering2013}, \citet{kennedy2014} and \citet{morales2016}, and so we search for any disagreement between these literature sources. In Table \ref{tab:target_properties} we list the literature references for each target, and specify which works find each target as having either one or two temperatures. We flag all those targets where there is disagreement in the literature as less certain. We then visually inspect the SEDs of targets for which there is only one literature source, and additionally flag the targets HD~120326 and HD~143675 as less certain. In both of these cases, no infrared excess is detected beyond the wavelength of the \textit{Spitzer} InfraRed Spectrograph (5.2-38\,$\micron$), and so it is hard to robustly infer a two-temperature disk. Our final target list includes 14 targets that host two-temperature debris disks, and 10 targets that likely host two-temperature debris disks, where this debris structure is less certain. All of the targets we observe are presented as having two temperatures in \citet{chen2014}, and so for consistency we use the temperature fits of that work in our subsequent analysis, with further details given in Section \ref{sec:diskradii}.

The final target list consists of 24 stars with some evidence for the presence of two belts. As part of the selection criteria, we included only stars with high parallaxes and young ages, since these targets allow the detection of planets at the closest physical separations to their host star, and at the lowest masses. The nearest OB2 association, Scorpius-Centaurus \citep[][hereafter Sco-Cen]{dezeeuw1999} is a particularly promising region for these studies since it is close ($\sim$140pc) and young ($\sim$10-16Myr, \citealt{pecautmamajek2016}). A significant fraction (58\%) of our targets are selected from this region. All of our targets have indicators of youth, mostly based on their association memberships, as detailed in Section \ref{sec:age}. Target properties are listed in Table \ref{tab:target_properties}.

\begin{table*}
	\centering
	\caption{Target stars. Distances are from \textit{Gaia} \citep{gaia_dr1b, gaia_dr1a} where available, and Hipparcos \citep{hipparcos} otherwise. USco, UCL, LCC indicate the Upper Scorpius, Upper Centaurus-Lupus, and Lower Centaurus-Crux regions of Sco-Cen respectively, while LA is the local association. Age determination is discussed in Section \ref{sec:age}.} 
	\label{tab:target_properties}
	\begin{tabular}{l l H H c c c c c}
		\hline
		HD & HIP & RA & Dec & Parallax[mas] & $\upsigma_{\textrm{para}}$[mas] & Association & Age[Myr] & Refs \\
		\hline
		\hline 
		166    & 544    & 00:06:36.785 & +29:01:17.40 & 72.63$^a$ & 0.52 & TWA/LA/Her-Lyr$^c$   & 8-150 & 6, 8, 10, 15 \\
                16743  & 12361  & 02:39:07.563 & -52:56:05.31 & 17.24$^a$ & 0.24 & Field    &  200 & 14 \\
		71722  & 41373  & 08:26:25.206 & -52:48:26.99 & 14.93$^a$ & 0.31 & Field    & 301$^{+227}_{-100}$ & 1, 2, 3, 11 \\
		79108  & 45167  & 09:12:12.881 & +03:52:01.11 & 10.07$^b$ & 0.39 & Field    & 212$^{+133}_{-67}$  & 1, 2, 3, 5 \\
		112810 & 63439  & 12:59:59.883 & -50:23:22.49 & 7.43$^a$  & 0.26 & LCC      & 17$\pm$1 & 4, 13 \\
		120326 & 67497  & 13:49:54.503 & -50:14:23.87 & 8.82$^a$  & 0.98 & UCL      & 16$\pm$1 & 4, 13 \\
                125541 & 70149  & 14:21:11.548 & -41:42:24.90 & 6.18$^a$  & 0.24 & UCL      & 16$\pm$1 & 4, 13 \\
		126062 & 70441  & 14:24:36.999 & -47:10:39.86 & 7.15$^a$  & 0.27 & UCL      & 16$\pm$1 & 4, 13 \\
                126135 & 70455  & 14:24:43.916 & -40:45:18.59 & 6.06$^b$  & 0.60 & UCL      & 16$\pm$1 & 4, 13 \\
		129590 & 72070  & 14:44:30.964 & -39:59:20.61 & 7.07$^a$  & 0.33 & UCL      & 16$\pm$1 & 4, 13 \\
                132238 & 73341  & 14:59:13.925 & -37:52:52.43 & 6.15$^b$  & 0.51 & UCL      & 16$\pm$1 & 4, 13 \\
		136246 & 75077  & 15:20:31.419 & -28:17:13.58 & 8.67$^a$  & 0.41 & UCL      & 16$\pm$1 & 4, 13 \\
		136482 & 75210  & 15:22:11.256 & -37:38:08.25 & 7.34$^b$  & 0.51 & UCL      & 16$\pm$1 & 4, 13 \\
		138965 & 76736  & 15:40:11.556 & -70:13:40.38 & 12.53$^a$ & 0.40 & Field    & 348$^{+39}_{-54}$ & 1, 3, 11 \\
                143675 & 78641  & 16:03:13.541 & -35:17:14.97 & 8.12$^a$  & 0.42 & UCL      & 16$\pm$1 & 4, 13 \\
		146606 & 79878  & 16:18:16.160 & -28:02:30.15 & 7.70$^a$  & 0.54 & USco     & 13$\pm$1 & 4, 13 \\
                148657 & 80897  & 16:31:11.673 & -38:22:58.74 & 6.04$^b$  & 1.15 & UCL      & 16$\pm$1 & 4, 13 \\
                151109 & 82154  & 16:47:01.679 & -39:32:01.94 & 4.96$^a$  & 0.91 & UCL      & 16$\pm$1 & 4, 13 \\
		153053 & 83187  & 17:00:06.279 & -54:35:49.84 & 19.30$^b$ & 0.35 & Field    & 539$^{+276}_{-268}$ & 1, 2, 3 \\
                182919 & 95560  & 19:26:13.246 & +20:05:51.84 & 13.72$^b$ & 0.34 & Field    & 198 & 2, 16 \\
		196544 & 101800 & 20:37:49.119 & +11:22:39.64 & 17.26$^b$ & 0.35 & Field    & 280$^{+256}_{-98}$ & 1, 3, 11 \\
		215766 & 112542 & 22:47:42.769 & -14:03:23.14 & 10.27$^b$ & 0.46 & Field    & 73$^{+115}_{-33}$ & 1, 2, 3 \\
                223352 & 117452 & 23:48:55.547 & -28:07:48.97 & 23.73$^b$ & 0.22 & AB Dor   & 150$^{+50}_{-30}$ & 9, 17 \\  %a, b, c
                225200 & 345    & 00:04:20.317 & -29:16:07.75 & 8.01$^b$  & 0.46 & Blanco I & 90$\pm$25  & 7, 12 \\
		\hline 
	\end{tabular}

	\begin{minipage}{\textwidth}
    \textbf{Note.} $^a$ \textit{Gaia} distance, $^b$ Hipparcos distance, $^c$ There is conflicting literature for this target, discussed in Section \ref{sec:age}. 
	
	\textbf{References.}  (1) \cite{brandt2015}; (2) \cite{chen2014}; (3) \cite{davidhillenbrand2015}; (4) \cite{dezeeuw1999}; (5) \cite{gerbaldi1999}; (6) \cite{lopezsantiago2006};  (7) \cite{lynga1984}; (8) \cite{maldonado2010}; (9) \cite{mamajek2016}; (10) \cite{nakajima-morino2012}; (11) \cite{nielsen2013};  (12) \cite{panagi1997}; (13) \cite{pecaut2012}; (14) \cite{rhee07}; (15) \cite{tetzlaff2011}; (16) \cite{zorecroyer2012}; (17) \cite{zuckerman2011}.
	\end{minipage}
	
\end{table*}

\begin{table}
	\centering
	\caption{Literature SED fits of each target as either one or two temperature disks. Since \citet{morales2011} and \citet{morales2016} use similar methodology we do not count these as independent, but use \citet{morales2016} where available and \citet{morales2011} otherwise. Note also that \citet{kennedy2014} do not present any one-temperature SED fits. In the final column, we list the targets for which we consider the two-belt nature to be more uncertain.}
	\label{tab:two-belt-fits}
	\begin{tabular}{l c c c}
		\hline
		HD & One-Temp & Two-Temp & Uncertain? \\
		\hline
		\hline 
		HD~166    &  & 1, 2, 5 \\
                HD~16743  &  & 2, 3 \\
		HD~71722  &   & 1, 2, 3, 5 \\
		HD~79108  &  & 1, 2, 3, 5 \\
		HD~112810 &  & 1, 2 \\
		HD~120326 & 1 & 2 & yes \\
                HD~125541 &  & 2 & yes \\
		HD~126062 &  & 2 & \\
                HD~126135 & 1, 5 & 2 & yes \\
		HD~129590 & 1 & 2 & yes \\
                HD~132238 &  5 & 2 & yes \\
		HD~136246 &  & 1, 2, 3, 5 \\
		HD~136482 &  & 1, 2, 3, 5 \\
		HD~138965 &  & 1, 2, 3, 5 \\
                HD~143675 & 1 & 2 & yes \\
		HD~146606 &  & 2 \\
                HD~148657 &  & 2 & yes \\
                HD~151109 &  & 2 \\
		HD~153053 &  & 1, 2, 3, 5 \\
                HD~182919 & 1, 4 & 2, 3 & yes \\
		HD~196544 &  & 1, 2, 5 \\
		HD~215766 &  5 & 1, 2 & yes \\
                HD~223352 &  5 & 1, 2 & yes \\  
                HD~225200 &  & 1, 2, 3, 4 \\
		\hline 
	\end{tabular}

	\begin{minipage}{\columnwidth}	
	\textbf{References.}  (1) \cite{ballering2013}; (2) \cite{chen2014}; (3) \cite{kennedy2014}; (4) \cite{morales2011}; (5) \cite{morales2016}.
	\end{minipage}
\end{table}

\section{Observations} \label{obs}

Each of the targets was observed with the SPHERE planet-finding instrument on the VLT \citep{beuzit2008}. Data were collected in the dual imaging IRDIFS mode, which splits the light into two subsystems: a differential imager and spectrograph \citep[IRDIS;][]{dohlen2008}, and an integral field spectrometer \citep[IFS;][]{claudi2008}. For this work we used IRDIS in dual-band imaging mode \citep[DBI;][]{vigan2010} with the H23 filter pair ($\lambda$\,=\,1588.8\,nm, $\Delta \lambda$\,=\,53.1\,nm and $\lambda$\,=\,1667.1\,nm, $\Delta \lambda$\,=\,55.6\,nm), and the IFS was used in the YJ mode, which spans the range 0.95-1.35\,$\micron$ and has 39 distinct wavelength channels \citep{zurlo2014,mesa2016}. Plate scales are 12.255~mas/pix for IRDIS and 7.46~mas/pix for the IFS \citep{maire2016}, and we use the \texttt{N\_ALC\_YJH\_S} coronagraphic mask, which has an inner working angle of $\sim$0.15$^{\prime\prime}$.

Each target was initially observed for a total integration time of $\sim$2000s, split into individual exposures between 2s and 64s. The individual exposure times were tailored based on the brightness and zenith distance of the target stars. The observations were carried out in pupil-stabilized mode to allow angular differential imaging \citep[ADI;][]{marois2006} to be performed. We also collected flux calibration frames, with the coronagraph removed, and star position calibration frames (waffle frames), where a sinusoidal pattern is applied to the deformable mirror to create four starspot images, one in each corner of the image. These allow the stellar position to be accurately measured behind the occulting mask. Flux and center calibrations were collected for each target, immediately before or after the main science observations.

For a subset of our target stars, follow-up observations were collected. These allow differentiation between background stars and co-moving companions based on whether the candidate shows common proper motion with the host. Follow-up observations generally had shorter exposure times, tailored to the specific candidates we were aiming to re-detect. Details of all observations (both initial and follow-up) used in this work are given in Table \ref{tab:obs}.
 
\begin{table*}
\centering
    \caption{SPHERE observations of target stars. The rotation column indicates the total rotation of the field, between the first and the last images. Note that the listed exposure times refer to each individual science image in the observation sequence.}
    \label{tab:obs}
    \begin{tabular}{ll|llHl|llHl}
        \hline
         & & \multicolumn{4}{c|}{IRDIS} & \multicolumn{4}{c}{IFS} \\
        Target & UT Date & $N_{\textrm{images}}$ & Exp. Time[s] & (red) & Rot[deg] & $N_{\textrm{images}}$ & Exp Time[s] & (red) & Rot[deg] \\
        \hline
        \hline
        HD 166    & 2015 Jul 19 & 768 &  2 & 11.164 & 11.2 & 180 &  4 &  8.907 & 8.9  \\
        HD 16743  & 2016 Sep 19 &  64 & 32 & 17.802 & 17.8 &  63 & 32 & 17.800 & 17.8 \\
        HD 71722  & 2015 Apr 25 & 192 &  8 & 15.061 & 15.1 &  82 & 16 & 13.082 & 13.1 \\
        HD 79108  & 2015 Apr 09 & 240 &  8 & 19.242 & 19.2 &  94 & 16 & 18.503 & 18.5 \\
        HD 112810 & 2016 May 02 &  80 & 32 & 21.763 & 21.8 &  40 & 64 & 22.879 & 22.9 \\
        HD 120326 & 2016 Jun 04 &  80 & 32 & 21.586 & 21.6 &  40 & 64 & 22.593 & 22.6 \\
        HD 125541 & 2015 Apr 16 &  64 & 32 & 29.925 & 29.9 &  62 & 32 & 29.682 & 29.7 \\
                  & 2016 Jun 04 &  16 & 32 &  6.871 & 6.9  &   8 & 64 &  6.688 & 6.7  \\
        HD 126062 & 2016 Jul 23 &  80 & 32 & 23.873 & 23.9 &  40 & 64 & 25.180 & 25.2 \\
        HD 126135 & 2016 Apr 07 & 512 &  4 & 34.027 & 34.0 & 448 &  4 & 35.157 & 35.2 \\
                  & 2018 Mar 17 &  64 & 32 & 24.778 & 24.8 &  32 & 64 & 26.532 & 26.5 \\
        HD 129590 & 2016 May 04 &  80 & 32 & 35.137 & 35.1 &  40 & 64 & 36.895 & 36.9 \\
        HD 132238 & 2016 Apr 07 & 128 & 16 & 29.139 & 29.1 & 124 & 16 & 29.263 & 29.3 \\
                  & 2018 Mar 17 &  64 &  8 &  7.616 &  7.6 &  17 & 32 &  8.076 &  8.1  \\
        HD 136246 & 2015 Apr 14 & 192 &  8 & 88.302 & 88.3 & 228 &  8 & 97.753 & 97.8 \\
                  & 2016 Apr 03 & 256 &  4 & 25.720 & 25.7 & 228 &  4 & 26.827 & 26.8 \\
        HD 136482 & 2015 Apr 15 &  96 & 16 & 35.073 & 35.1 &  67 & 16 & 24.925 & 24.9 \\
        HD 138965 & 2015 Apr 15 & 240 &  8 & 12.209 & 12.2 &  31 & 32 &  6.231 & 6.2  \\
        HD 143675 & 2015 Jul 11 &  96 & 16 & 36.837 & 36.8 &  91 & 16 & 35.596 & 35.6 \\
        HD 146606 & 2016 Jul 02 & 304 &  8 & 15.767 & 15.8 & 150 & 16 & 17.665 & 17.7 \\
        HD 148657 & 2015 Apr 20 &  96 & 16 & 34.510 & 34.5 &  89 & 16 & 33.280 & 33.3 \\
                  & 2016 Jun 05 &  16 & 32 &  8.581 & 8.6  &   8 & 64 &  8.388 & 8.4  \\
        HD 151109 & 2015 Apr 15 &  96 & 16 & 31.943 & 31.9 &  91 & 16 & 30.765 & 30.8 \\
                  & 2016 Jun 04 &  16 & 32 &  8.041 & 8.0  &   8 & 64 &  7.857 & 7.9  \\
        HD 153053 & 2015 Apr 23 &  96 & 16 & 13.611 & 13.6 &  45 & 32 & 12.958 & 13.0 \\
                  & 2016 Apr 09 &  64 & 32 & 15.438 & 15.4 &  64 & 32 & 15.676 & 15.7 \\
        HD 182919 & 2016 Apr 14 & 128 & 16 & 12.059 & 12.1 &  64 & 32 & 12.212 & 12.2 \\
                  & 2017 Jul 15 &  48 & 32 &  8.273 &  8.3 &  24 & 64 &  8.883 &  8.9 \\
        HD 196544 & 2015 May 29 &  48 & 32 & 10.980 & 11.0 &  69 &  8 & 13.743 & 13.7 \\
        HD 215766 & 2015 Jun 20 & 192 &  8 & 28.143 & 28.1 & 228 &  8 & 37.978 & 38.0 \\
        HD 223352 & 2015 Jul 16 & 320 &  4 & 65.063 & 65.1 & 196 &  8 & 82.852 & 82.9 \\
        HD 225200 & 2015 Jul 18 &  64 & 24 & 31.304 & 31.3 &  60 & 32 & 35.276 & 35.3 \\
        \hline
    \end{tabular}
\end{table*}

\section{Data Reduction} \label{datared}

\subsection{Pre-processing}

\subsubsection{IRDIS}

Pre-processing of the SPHERE/IRDIS data was performed using the CPL (Common Pipeline Library) provided by ESO. Master dark and flat frames were created, and the star position behind the coronagraph was calibrated using the center calibration frames. Each data frame was independently reduced by applying the master dark and flat frames, and then realigned taking into account the star center position calibration and the dither position for each frame.

\subsubsection{IFS}

Integral Field Spectrograph (IFS) data reduction was performed following \citet{vigan2015sirius}. Basic calibrations were first created using the ESO data reduction and handling pipeline \citep[DRH,][]{pavlov2008}: master dark and flat fields, IFS spectral position calibrations, initial wavelength calibrations and an IFU flat-field were all created. We then used a custom pipeline to calculate accurate time and parallactic angles for each image, and to normalise the data based on the direct integration time and neutral density filters for each observation. The pipeline also performs bad pixel correction and cross-talk correction, and the DRH is then used to interpolate these frames spectrally and spatially. To complete the initial cleaning and calibration of the frames, we finally perform a sigma-clipping routine to remove remaining bad pixels, and a correction of the wavelength calibration. Full details of these cleaning and calibration steps are given in \citet{vigan2015sirius}.

\subsection{Principal Component Analysis}

After the initial cleaning and calibration of the data, we use Principal Component Analysis \citep[PCA, see e.g.][our own implementation]{soummer2012,amaraquanz2012} to remove stellar speckle noise. The same process is carried out on both the IRDIS and the IFS data. We perform a full-frame PCA, taking into account each timestep (typically $\sim$90) and each wavelength channel (2 for IRDIS data, 39 for the IFS) independently. First, each wavelength channel is rescaled proportional to its wavelength, such that the characteric scale of speckles is equal between images. Then, the PCA algorithm is applied to remove stellar speckles based on the similarities between each individual image: speckles appear at the same location in each scaled image, while on-sky signals (planets, debris disks or background stars) appear at different positions with time since the field is rotating, and with wavelength due to the image scaling. The PCA processed images are then rescaled back to their original plate scales, and the parallactic angle for each image is used to align the North axis of each time step. The individual images are finally co-added to give a single, broadband reduced image for each target. 
Following the same process but co-adding by time only, we also create a cube of reduced images at each individual wavelength. This allows a comparison of H2 and H3 magnitudes in the case of IRDIS data, and spectral extraction across the YJ bands in the case of IFS data. 

The aggressiveness of the PCA algorithm is tuneable: removing more principal components before co-adding the images removes more of the scattered starlight, but also reduces the throughput of the planetary signal. We aim to achieve the optimum balance between removing starlight and preserving companion signal, so as to detect the faintest possible planets and place the most stringent contrast limits. To do this, we perform several PCA reductions with the same code, where we remove between 1 principal component and approximately one-third of the total available principal components, at which point a planetary signal is almost entirely removed. Each of these different reductions is used in our subsequent analysis when identifying candidate companions and calculating contrast curves.

\begin{figure*}
    \includegraphics[width=0.9\textwidth, trim=0cm 0cm 7cm 0cm]{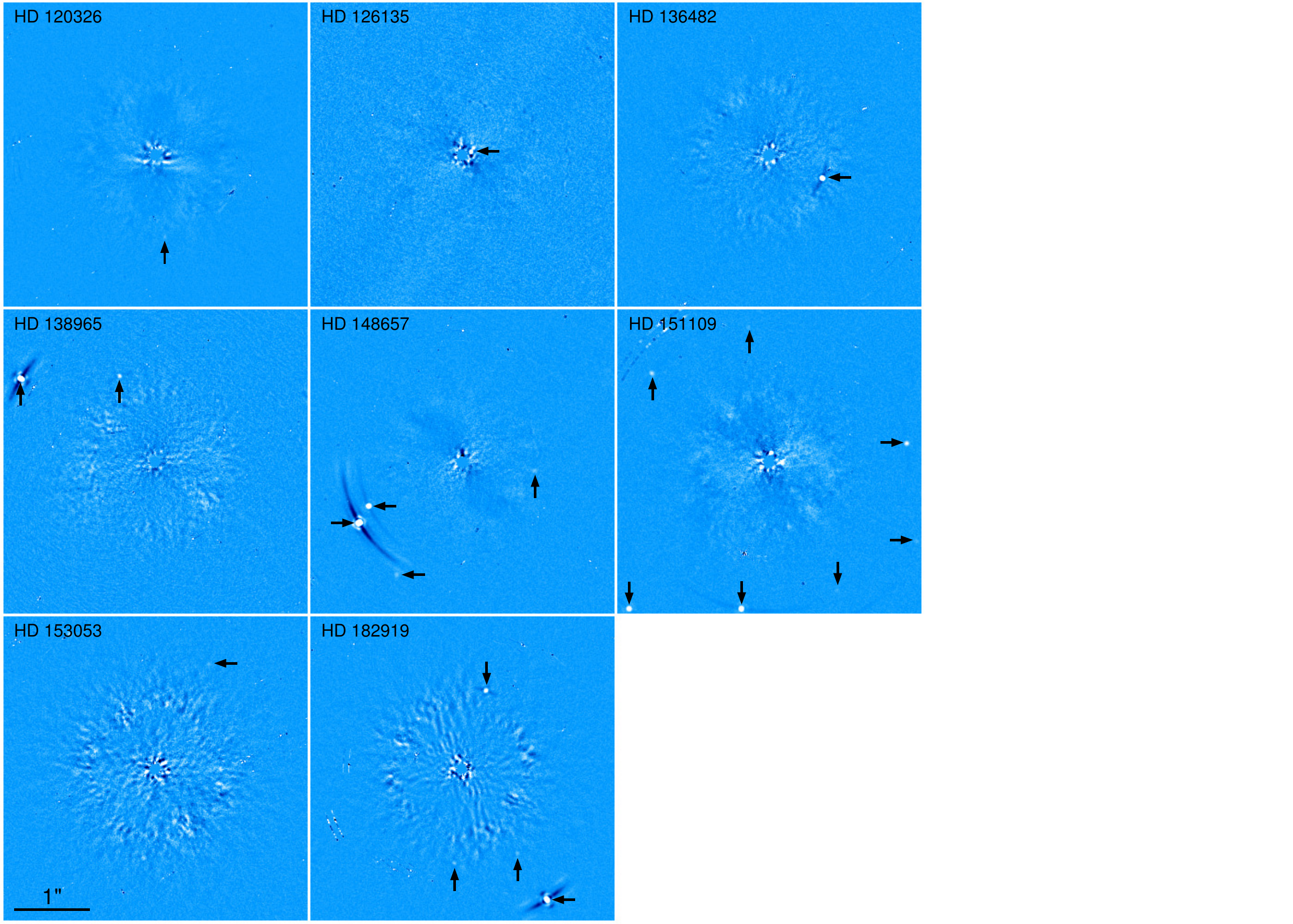}
    \caption{PCA reduced images for a selection of targets from the survey. All survey targets with companions closer than 2'' are shown. Each image is 4'' square, and candidates within this field of view are highlighted with arrows. The HD~126135 candidate is a likely speckle, as detailed in Section \ref{sec:126135}. The arcsecond scale bar applies to all images, the colorbar is identical for each thumbnail, and North is oriented upwards in each case.}
    \label{fig:thumbnails}
\end{figure*}

\subsection{Candidate Companion Identification \& Verification}

Candidate companions were identified by visual inspection of both the IRDIS and the IFS data, and each target was visually inspected by at least two individuals to confirm that no candidates were missed. This process is repeated at several PCA reduction strengths, as described above. Since SPHERE data are very high quality and most of the candidates are in the wide field where the data is read-noise limited, we include all visually identified candidates in our analysis rather than applying a sigma cut-off. Candidates within a 2'' square, centred on the host star, are shown in Figure \ref{fig:thumbnails}. 

IRDIS observations are used to calculate astrometry of each candidate companion relative to its host star. To do this, the pixel position of each candidate and the stellar position behind the occulting mask are measured, and we assume an error of 0.2 pixels in the determination of each. Following the ESO User Manual\footnote{6th \& 7th release, see https://www.eso.org/sci/facilities/paranal/ instruments/sphere/doc.html}, we use a plate scale of 12.255$\pm$0.021mas/pix and a true north correction of -1.700$\pm$0.076$^{\circ}$ for data before December 2015, and \hbox{-1.75$\pm$0.08$^{\circ}$} for data after February 2016. The additional pupil IFS offset (135.99$\pm$0.11$^{\circ}$) is also applied, as well as an additional `epsilon' correction, due to a missynchronisation problem at the telescope (see the User Manual for details). For our data, we find that this correction is consistently smaller than 0.1$^{\circ}$. We also correct for the anamorphic distortion of the chip before performing astrometry. The epsilon and anamorphic distortion corrections are applied to each individual frame, before the images are combined. The measured separation and position angle of each candidate are listed in Table \ref{tab:candidates}. 

We use several methods to distinguish between genuine companions and background objects: we refer to previous literature, use common proper motion testing where there are multiple epochs of SPHERE data, and study the H2-H3 colors for candidates with an absolute magnitude fainter than 15 in the H2 filter. A negative H2-H3 color indicated the presence of methane, which we expect for sufficiently low mass companions but not for distant background stars. For the remaining candidates, it is not possible to make a conclusive determination, but we use separation from the host star to determine likely background objects. In this survey we detect 178 candidates of which 2 have been previously published as companions, and 13 have been previously published as background objects. A further 124 are found to be background objects based on their common proper motion between two epochs, and 20 are background objects based on their H2-H3 colors. The final 18 are likely background objects based on their relative faintness and wide separations from their respective host stars. The final designation of each candidate is given in Table \ref{tab:candidates}. One additional candidate to HD~126135 is detected close to the coronagraph in a first epoch of data, but not redetected and we conclude the object is likely to be a speckle.
\label{sec:candidatedesignation}

\begin{figure*}
    \includegraphics[width=\textwidth]{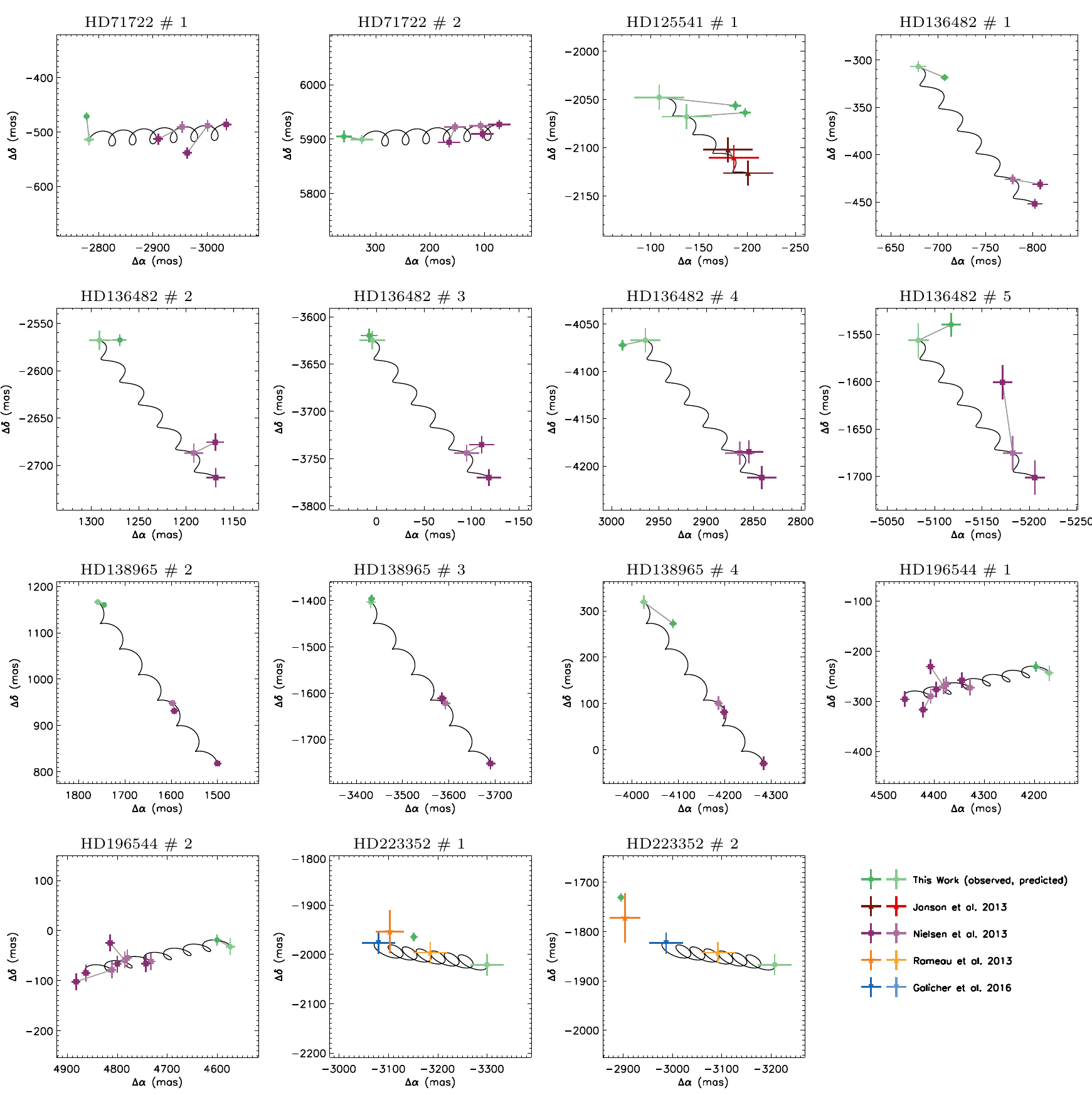}
    \caption{Astrometry for all candidates with archival data. Archival data are taken from \citet{nielsen2013,janson2013a,rameau2013,galicher2016}. Darker points are measured, and lighter points show the predicted position for a background object at each epoch, with the black lines showing the path a stationary background object would take. In several cases there is imperfect agreement with the background hypothesis, possibly due to non-zero motion of the background objects. Although we see a systematic offset between \citet{janson2013a} and our astrometry for HD~125541, discussed further in Section \ref{sec:jansondifference}, we agree with their conclusion that this is a background object. For HD~223352, we only plot a subset of archival astrometry for clarity, and the two candidates are previously confirmed companions, as discussed in Section \ref{sec:indietargets} and Table \ref{tab:223352}.}
    \label{fig:gridfig_comparative}
\end{figure*}

For the 13 candidate companions that have been previously determined to be background objects, we plot  relative astrometry against the published astrometry in Figure \ref{fig:gridfig_comparative}. We consistently see close agreement with the predicted positions for candidates in \citet{nielsen2013}, and in each case confirm their conclusion that these are background objects. For the candidate around HD~125541 that was previously published in \citet{janson2013a}, we note a systematic offset of $\sim$80mas in the candidate astrometry between their work and our measurements. It is not immediately clear what the cause of this difference is, but since the candidate is relatively bright ($\Delta$H2=8.6mag) we suggest that it is a non-infinite background object with non-zero proper motion. The 7 candidate companions to HD~120326 were previously detected in \citet{bonnefoy2017}, but due to the short time baseline between their observations and ours we do not attempt to create CPM plots for this target. Two candidate companions to HD~223352 have been previously detected in several works \citep{derosa2011,rameau2013,galicher2016}, and confirmed to be co-moving. These companions are discussed in more detail in Section \ref{sec:223352}.
\label{sec:jansondifference}

\begin{figure*}
    \includegraphics[width=\textwidth]{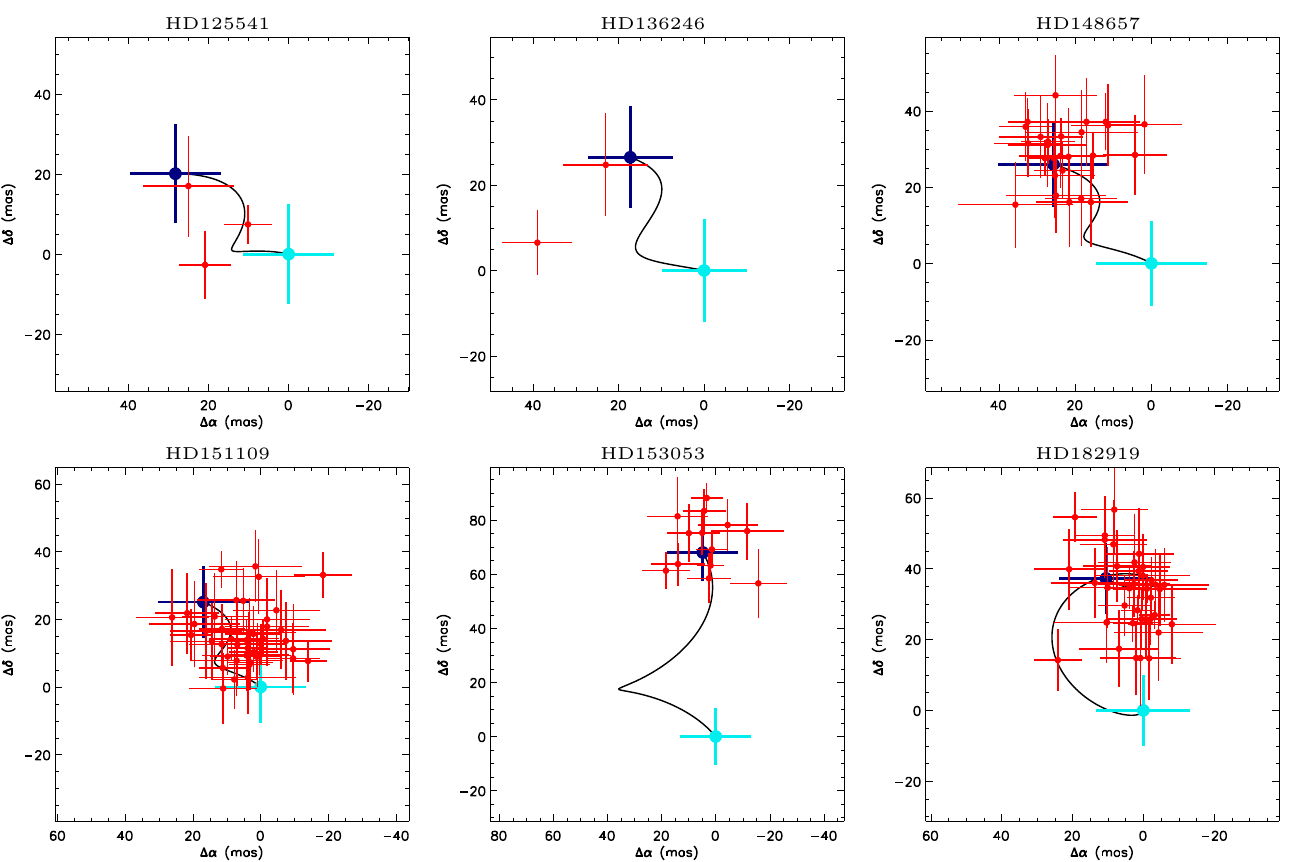}
    \caption{Multi-candidate common proper motion plots for targets where we have multiple epochs of SPHERE data. The predicted background motion for a candidate in each case follows the black line from the light to the dark blue point, and the measured final positions of each candidate are plotted in red relative to the light blue point. The complete astrometry is included in Table \ref{tab:candidates} and individual common proper motion plots for each candidate are included in Appendix \ref{app:cpm}.}
    \label{fig:gridfig_alls}
\end{figure*}

For 6 targets, we have multiple epochs of SPHERE data. In these cases we create multi-candidate common proper motion plots (see Figure \ref{fig:gridfig_alls}), where the motion of each candidate relative to its host is presented simultaneously. A reference track, demonstrating the predicted motion of an infinitely distant background star relative to the primary is also plotted. HD~148657 and HD~153053 demonstrate the expected outcome for a target with a large number of candidate companions: the final positions of the various candidates (in red) are clustered around the predicted final position (dark blue), with some statistical spread. In these cases it is clear that each of the plotted companions shows a good match to the background hypothesis. In the case of HD~151109, however, the measured final positions of candidates are clustered around a point in between the initial (light blue) and predicted final (dark blue) positions. This is indicative of some systematic error: either (a) the host star position is incorrectly calibrated behind the coronagraphic mask, (b) there is a slight error in calibration of the telescope angle or (c) the proper motion and parallax of this object in the \textit{Gaia} catalog \citep{gaia_dr1b,gaia_dr1a} are not accurate. This systematic uncertainty can be probed by considering the candidates simultaneously, since \textit{all} the candidates show this shift from the expected final position, and it is clear that not all the candidates are genuine companions. We suggest instead that any candidates with significantly outlying proper motion relative to the other candidates should be considered as co-moving companions, rather than any candidates which show a small proper motion between the two epochs. For HD~151109, therefore, all the candidates appear to be background objects. For a subset of targets with faint enough $M_{H2}$ for the H2-H3 colors to differentiate between companions and background objects, we find H2-H3 colors close to zero, further supporting this conclusion. In cases like this the entire set of candidates reveals additional information about systematics: although an individual CPM diagram might suggest a co-moving companion, comparing the entire set of candidate in this way allows more accurate conclusion to be drawn about the true nature of candidates.
\label{multi-astrometry-text}

\begin{table*}
 	\centering
 	\caption{Candidate companion astrometry and magnitudes for the survey. A total of 178 candidates were detected, of which 157 are background (BG) and a further 18 are likely background objects (?BG) based on their separation from the host star and color analysis. 2 objects are previously detected companions (C) and one object is a likely speckle (S?). Further detail on candidate designation is given in Section \ref{sec:candidatedesignation}. \textit{Only a portion of the table is shown here - the entirety will be included in the online journal as supplementary material.}}
 	\label{tab:candidates}
 	\begin{tabular}{lcHHcccccccc} 
 		\hline
 	    Star & Epoch Date & No./DEW & No./ECM & No. & $\Delta$H (mag) & Sep('') & $\upsigma_{\textrm{sep}}$ & PA & $\upsigma_{\textrm{PA}}$ & Reference & Status \\
 	    \hline
 	    \hline
        HD 71722	&	2015-04-25	&	1	&	1	&	1	&	11.4	&	2818.2	&	6.0	&	260.37	&	0.15	&	N13	&	BG	\\
        HD 71722	&	2015-04-25	&	2	&	2	&	2	&	11.2	&	5915.2	&	10.8	&	3.47	&	0.14	&	N13	&	BG	\\
        \hline
        HD 79108	&	2015-04-09	&	1	&	1	&	1	&	13.3	&	5248.5	&	9.7	&	89.50	&	0.14	&	--	&	?BG	\\
        \hline
        HD 122810	&	2016-05-02	&	2	&	1	&	1	&	14.6	&	3512.9	&	7.0	&	262.04	&	0.15	&	--	&	?BG	\\
        HD 122810	&	2016-05-02	&	4	&	3	&	2	&	7.8	&	5682.9	&	10.4	&	311.36	&	0.14	&	--	&	?BG	\\
        HD 122810	&	2016-05-02	&	5	&	4	&	3	&	11.3	&	5788.2	&	10.6	&	357.93	&	0.14	&	--	&	?BG	\\
        HD 122810	&	2016-05-02	&	1	&	5	&	4	&	13.8	&	5859.7	&	10.7	&	126.75	&	0.14	&	--	&	?BG	\\
        HD 122810	&	2016-05-02	&	3	&	2	&	5	&	--	&	6188.0	&	11.2	&	284.11	&	0.14	&	--	&	?BG	\\
        \hline
        \hline
 	\end{tabular}
 	
 	\vspace{0.2cm}
 	
 	\begin{minipage}{\textwidth}
 	\vspace{0.1cm}
    \textbf{Note.} TF = candidate too faint to be redetected, OS = candidate off-screen in this epoch.
	
	\textbf{References.} (B17) \cite{bonnefoy2017}; (dR11) \cite{derosa2011}; (G16) \cite{galicher2016}; (J13) \cite{janson2013a}; (N13) \cite{nielsen2013}; (R13) \cite{rameau2013}.
	\end{minipage}
\end{table*}

\subsection{Contrast Limits}
\label{sec:contrasts}
For each of the targets, contrast limits are calculated via injection of fake candidates. Several scaled images of the PSF calibration frame are inserted into the raw data at a variety of offsets and position angles, and the full reduction process repeated. A total of 20 scaled PSF images are inserted into each IFS frame, and 60 into each IRDIS frame. In each case, the minimum separation between fake planets is 100mas, to avoid contamination between the separate injections. The injections are repeated at five different position angles, and at several different magnitudes. The contrast quoted in this work is the mean 5$\sigma$ detection across the five fake planet candidates at each separation. To account for the small number of resolution elements at small inner working angles, the correction term presented in \citet{mawet2014} is applied. By using this method we ensure the planetary throughput of the algorithm is accurately captured.

This process is performed for each of our PCA reductions with different numbers of PCA components removed, and the contrast quoted is that of the most favorable reduction. By testing the contrast at a variety of reduction strengths, we ensure that we remove the optimum number of PCA modes to balance removing sufficient starlight, while minimizing the extent to which the planetary signal is self-subtracted for each individual dataset.

We convert these contrast limits into mass limits by using the COND models \citep{baraffe2003} for temperatures below 1700K and DUSTY models \citep{chabrier2000} otherwise \citep[as in e.g.][]{janson2013a}. For simplicity, we use only the SPHERE/IRDIS data in calculating these mass limits. The majority of conversions use the COND models, due to the high sensitivity of the SPHERE instrument.  

\section{Results} \label{results}

\subsection{Achieved contrast}

\begin{figure*}
	\includegraphics[width=1.8\columnwidth]{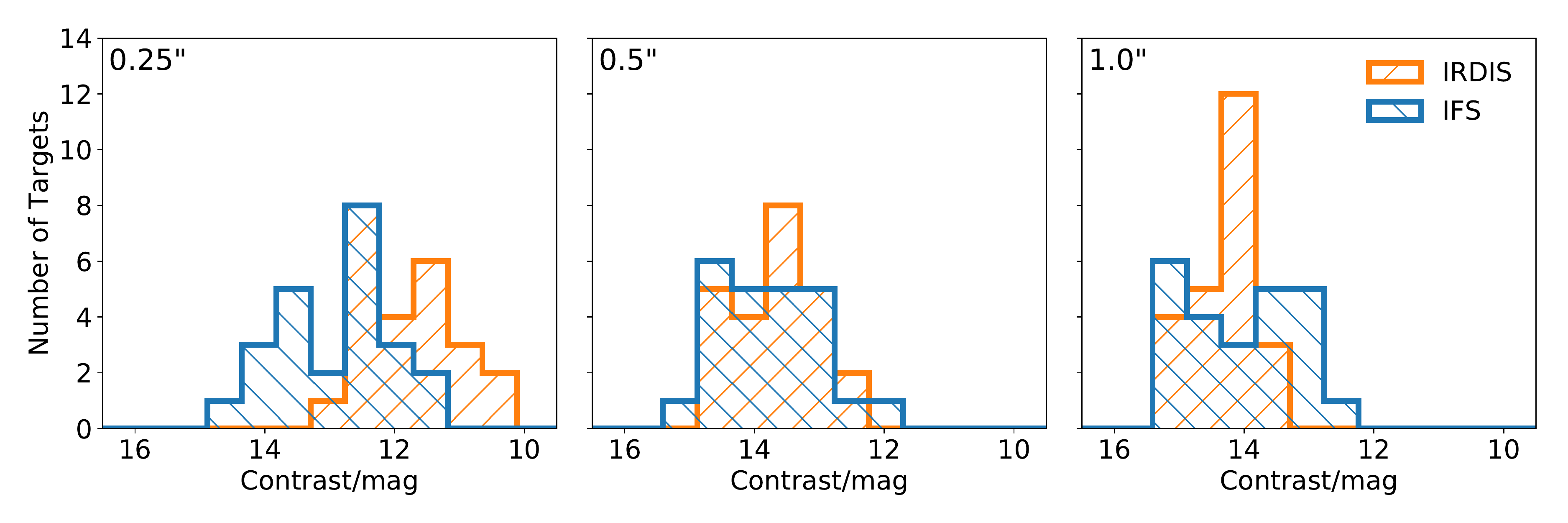}
    \caption{Histograms of our survey contrast, at a separation of 0.25'', 0.5'' and 1.0''. Orange and blue lines represent the IRDIS and IFS data respectively.}
    \label{fig:contrasthist}
\end{figure*}

Our achieved contrast as a function of separation for both the IRDIS and IFS instruments is presented in Figure \ref{fig:contrasthist}. Mean and best contrasts as a function of separation are given in Figure \ref{fig:meanbestcontrast} and individual contrast curves for each dataset are presented in Appendix \ref{app:contrasts}. We are able to reach contrasts of $\sim$15 magnitudes at 0.5$''$ in the most favorable systems.

\begin{figure}
	\includegraphics[width=1.\columnwidth]{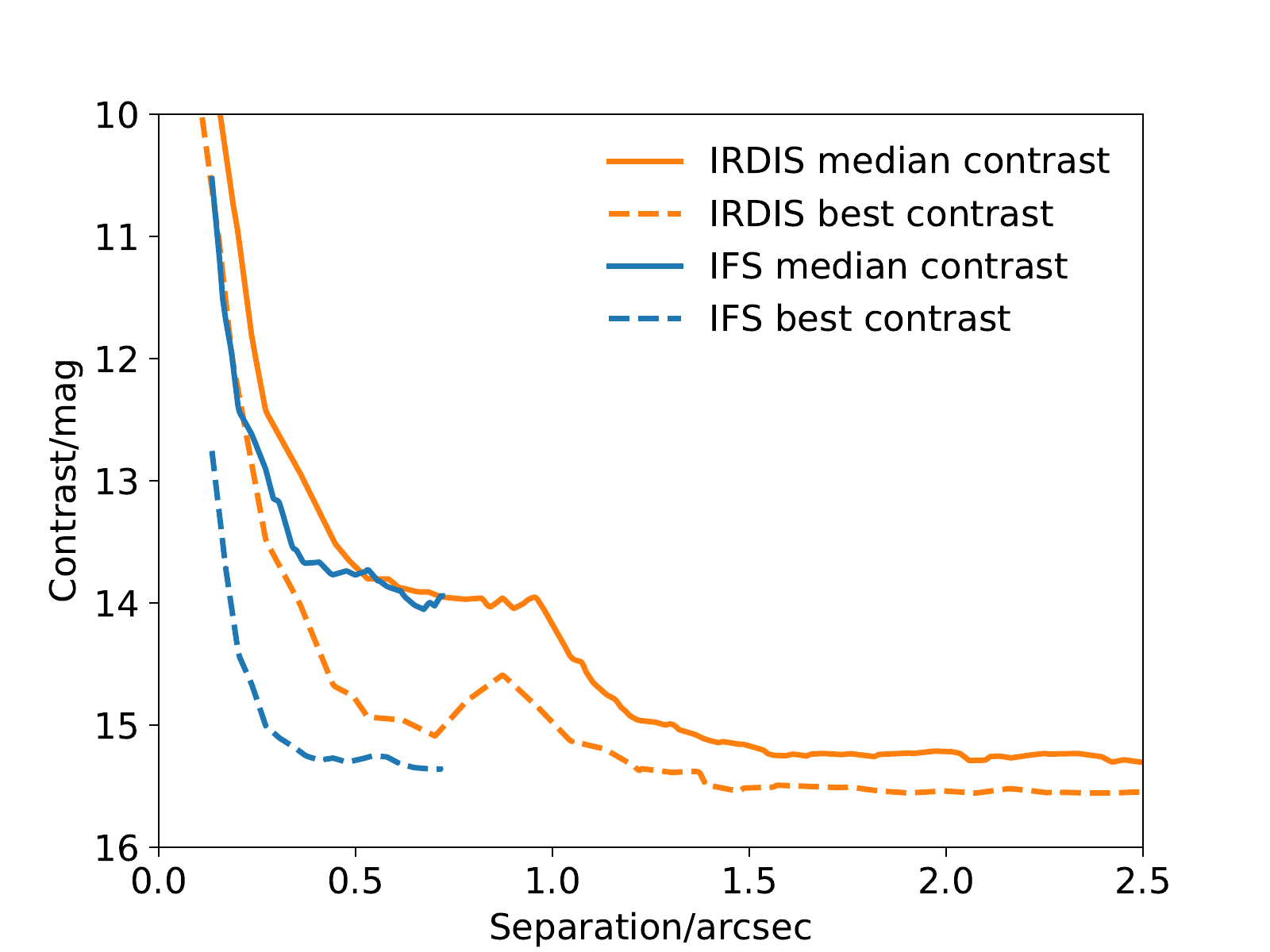}
    \caption{Median and best contrasts achieved by our survey, for both the IFS and IRDIS subsystems. Only the initial observation (durations $\sim$1h) of each target is included in this plot. }
    \label{fig:meanbestcontrast}
\end{figure}

\subsection{Disk Radii}
\label{sec:diskradii}

14 of the targets in this work show strong evidence for hosting two-temperature debris disks, based on the available literature and our examination of the SEDs as described in Section \ref{targets}. The remaining 10 targets are less certain: these target SEDs can be modelled almost as well with a single temperature excess as with two temperatures. For the purposes of this paper, we proceed under the assumption that these two-temperature systems host two debris belts and discuss the planetary configurations for this case. If these are in fact single debris belt systems, there are clearly a range of additional planetary configurations which are not considered in this work.

For consistency, we use the temperature values found in \citet{chen2014} to calculate radii for all of our targets. We calculate updated radii following \citet{pawellek2015} and using the ``50\% astrosilicate + 50\% ice" dust composition.

\subsection{Resolved Disk Radii}

For six of the targets in this survey, resolved disk images exist, and these targets are listed in Table \ref{tab:resolvedradii}. This allows some verification of the calculated radii. Four of the targets have been resolved with \textit{Herschel} \citep[see][]{morales2016}, and in two of these cases, namely HD~71722 and HD~138965, we see close agreement with the calculated values. For the other two targets, there is a factor $\sim$2 difference between the measured and calculated radii, which changes our calculated lower mass limits (see below) by a factor $\sim$2.8.

Two of the targets, namely HD~120326 and HD~129590, are resolved with VLT/SPHERE at $\sim$1.6$\micron$. For HD~129590, the resolved radius shows very close agreement with the calculated radius for the inner dust belt. Given that \citet{matthews2017} found a very soft external power law for the dust ring, we suggest that the resolved disk corresponds to the inner dust belt, and that the soft power law is caused by an additional, fainter ring of dust at wider separation. For HD~120326, \citet{bonnefoy2017} found evidence for both dust belts in scattered light, and both radii are listed below. The outer radius matches closely with the calculated value, and the inner radius is within a factor of 2. We note at this point that the calculated lower mass limits (see Section \ref{sec:shannon-lower}) depend only on the radius of the outer disk, although the position of the inner disk determines the number of planets at this mass that are required to fill the gap.

\begin{table*}
    \centering
    \caption{Measured and calculated radii for disks where at least one belt of debris has been resolved. The upper group of targets have been resolved with the \textit{Herschel} space telescope, and for these we list only the calculated outer radius, which corresponds well for two targets and is a factor of 2 off for two targets. The lower group have been resolved with VLT/SPHERE. In this case we list both calculated radii: for HD~129590 the resolved disk is likely the inner band of dust, while for HD~120326 both bands of dust are tentatively detected in \citet{bonnefoy2017}, with the outer closely matching the calculated value. Note that only the outer disk radius is used in the calculation of a lower mass limit.}
    \begin{tabular}{llccl}
        Target & $\lambda$ & Resolved Radius / AU & Calculated Radius / AU & Reference \\
        \hline \hline
        HD~166 & 70$\micron$ & 29$\pm$3 & $76^{+12}_{-10}$ & \citealt{morales2016} \\
        \vspace{0.4em}
               & 100$\micron$ & 36$\pm$3 &  & \citealt{morales2016} \\
        \vspace{0.4em}
        HD~71722 & 100$\micron$ & 139$\pm$27 & $128^{+20}_{-16}$ & \citealt{morales2016} \\
        \vspace{0.4em}
        HD~138965 & 100$\micron$ & 187$\pm$6 & $191^{+38}_{-30}$ & \citealt{morales2016} \\
        \vspace{0.4em}
        HD~153053 & 100$\micron$ & 186$\pm$12 & $306^{+81}_{-68}$ & \citealt{morales2016} \\
        \hline
        \vspace{0.4em}
        HD~120326 & 1.6$\micron$ & 58.6$\pm$3, 130$\pm$8 & $33.0^{+2.8}_{-2.4}$, $134^{+24}_{-19}$ & \citealt{bonnefoy2017} \\
        \vspace{0.4em}
        HD~129590 & 1.6$\micron$ & 59.3$\pm$0.2 & $60.2^{+1.3}_{-1.3}$, $103^{+451}_{-3}$ & \citealt{matthews2017} \\
        \hline
        \vspace{0.4em}
    \end{tabular}
    \label{tab:resolvedradii}
\end{table*}

\subsection{Age Determination}
\label{sec:age}
Where available, we use cluster membership to determine the ages of each target. 14 of the 24 targets are members of the Scorpius-Centaurus association, as determined by \citet{dezeeuw1999}, and we use the \citet{pecaut2012} ages for each Sco-Cen subgroup. There is some disagreement about the membership of HD~166: it is listed as a member of either Hercules-Lyra \citep[150-300Myr;][]{lopezsantiago2006}, the Local Association \citep[20-150Myr;][]{maldonado2010} or the TW Hydrae association \citep[8Myr;][]{nakajima-morino2012}. \citet{tetzlaff2011} also find a very young age of 20.1$\pm$6.4Myr for this target  using pre-main sequence evolutionary models. We choose to assign this target a range of ages, namely 8-150Myr, to reflect this range of literature ages, and in subsequent calculations represent this range as an age of 79$\pm$71Myr. HD~223352 is a member of the AB Dor moving group \citep{zuckerman2011}, which has an age of 150$^{+50}_{-30}$Myr \citep{mamajek2016}. HD~225200 is a member of Blanco I \citep{lynga1984}, which has an age of 90$\pm$25Myr \citep{panagi1997}.

The remaining targets are field stars, and so ages are harder to determine accurately. Each is nonetheless likely to be young, given the presence of high volumes of circumstellar dust. For these targets we use previously performed age determinations. HD~71722, HD~79108 and HD~196544 all show close agreement between several literature sources (see Table \ref{tab:target_properties}), and in these cases we use the Bayesian ages from \citet[][here on DH15]{davidhillenbrand2015}. For HD~138965 and  HD~215766, there is some slight discrepancy between DH15 and \citet{brandt2015}, with the best fit ages varying by a factor of $\sim$3. For consistency, we use the DH15 ages here too, but note that there is more uncertainty. For HD~153053, the DH15 Bayesian age appears discrepantly lower than both the DH15 interpolated age and the ages presented by \citet{brandt2015} and \citet{chen2014}, and so we use the Brandt age. Finally, there is limited literature for both HD~16743 and HD~182919 and so we use \citet{rhee07} and \citet{zorec2012} respectively, but note that these age designations are more uncertain. In these two cases, no uncertainties are quoted with the literature ages.

\subsection{Candidate Companions to Individual Targets}
\label{sec:indietargets}

\hspace{1em}\textit{HD~166:} This target was previously studied by \citet{lafreniere2007} as part of the GDPS, and no candidates were identified. Even with our improved contrast limits, we do not find any candidates around HD~166.

\hspace{-1em}\textit{HD~16743:} No candidates are identified around HD~16743.

\hspace{-1em}\textit{HD~71722:} Both the candidates presented in \citet{nielsen2013} as background stars are redetected in this work, and our astrometry is consistent with that of \citet{nielsen2013} for the background hypothesis. No further candidates are identified.

\hspace{-1em}\textit{HD~79108:} For this target, we identify a single candidate at a separation of 5.25'' and $M_{H2}$=14.5. This is too bright for H2-H3 color analysis to be conclusive, but the physical projected separation of 521~AU strongly suggests a background object.

\hspace{-1em}\textit{HD~112810:} Five widely separated candidates are identified (>3.5''). All are likely background objects based on their separation. Additionally, the disk was detected in scattered light for the first time (Matthews et al., in prep).

\hspace{-1em}\textit{HD~120326:} A debris disk was imaged around this target in \citet{bonnefoy2017}. We redetect this debris disk, and detect seven of the ten candidate companions found in that work. An eigth candidate appears on the very edge of the detector where astrometric measurements are no longer reliable, and we choose to ignore this candidate. The final two candidates listed in \citet{bonnefoy2017} are off the edge of our detector, due to the camera rotation. Our data were collected two months after those in \citet{bonnefoy2017}, a short time baseline in which a background object would move 11.2mas relative to the host star. Although this number is larger than the nominal SPHERE astrometric accuracy of 5mas, it is too small to allow us to clearly differentiate the companion and background hypotheses, and we do not create common proper motion plots for this target. \citet{bonnefoy2017} conclude that all of these candidates are background objects based on their colors, and on previous detections of several of the candidates in HST/STIS data \citep{padgett2016}.

\hspace{-1em}\textit{HD~125541:} This candidate was observed twice, with four candidates detected in the first epoch, and three of these redetected in the second epoch. Candidate \#1 was previously detected in \citet{janson2013a} and confirmed to be a background object. We detect significantly less than the expected proper motion between our two observational epochs. Given the systematic differences with \citet{janson2013a} and the relative brightness of the candidate, this is likely a nearby background object, with non-zero proper motion. Candidates \#2 and \#4 both show significant motion between our two observational epochs, suggesting that they are background objects. Candidate \#3 is only detected in one epoch, at $M_{H2}$=16.1. At this very faint magnitude, an H2-H3 color of 0.08 and a separation of 4.86''=786~AU imply that this is a background object.

\hspace{-1em}\textit{HD~126062:} Three faint, wide separation candidates are identified around this target. With only one epoch of data, we are unable to use proper motion to confirm whether the candidates are genuine companions or background objects. Based on the wide separation, faint absolute magnitude and low H2-H3 color of each candidate, all three are assumed to be background objects.

\hspace{-1em}\textit{HD~126135:} In a first epoch of data, we find a bright candidate very close to the coronagraph edge (separation 137mas, see Figure \ref{fig:thumbnails}). At this close separation it is hard to distinguish companions and speckle noise, but the candidate is resilient to the number of principal components subtracted, and appears to have self-subtraction wings. The candidate appears in the IRDIS but not the IFS data, suggesting that it is either an extremely red object or a speckle.

In a second epoch of data the candidate is not recovered. Although it is possible that this is a genuine low-mass companion, it is most likely a particularly persistent speckle, and further follow-up is required to confirm the nature of this object.
\label{sec:126135}

\hspace{-1em}\textit{HD~129590:} The debris disk around this target was detected in scattered light for the first time \citep[see][]{matthews2017}. In addition, one candidate was identified at 5.67'', corresponding to a physical projected separation of 752~AU. At this wide separation, the candidate has a low probability of being associated with the host star. The candidate is positioned North of the debris disk, which has a position angle of 122$^\circ$ and an inclination of 75$^\circ$ \citep{matthews2017}. A bound candidate in this position would either be significantly further than this 752~AU separation, or significantly misaligned with the disk, further supporting our assumption that this is a background star and not a bound companion.

\hspace{-1em}\textit{HD~132238:} A single candidate is observed at a separation of 4.29''. The candidate has $M_{H2}$=15.3 and H2-H3=0.05, and shows good agreement with the predicted motion of a background object between two epochs, and so we conclude that it is a background object.

\hspace{-1em}\textit{HD~136246:} Two candidates are identified, and both are redetected in a second epoch of data. Although the astrometric measurement of candidate \#1 is displaced from the predicted position in epoch 2, the candidate moves significantly from the initial position. Since a companion would show almost no motion relative to the host in this period, this is likely a background star with non-zero proper motion. As such, we conclude that both candidates are background stars.

\hspace{-1em}\textit{HD~136482:} Six candidates are identified around HD~136482. Five of these have been previously identified by \citet{nielsen2013}, and an additional candidate at 5.95'' is found below the contrast limit in that work. Based on the projected separation and H2-H3 color of this candidate, it is a background object. We do not detect the 6th candidate listed in \citet{nielsen2013} since it is outside the SPHERE field of view. 

\hspace{-1em}\textit{HD~138965:} Four candidates are detected, three of which are also listed in N13 as background objects. Our candidate \#1 is below the detection limit of N13, with $M_{H2}$=15.2. For this candidate H2-H3=0.05, and so the candidate is a likely background object. 

\hspace{-1em}\textit{HD~143675:} 4 candidates are detected around HD~143675. Since all are faint (contrast 11.2 mag or higher) and at wide separation ($>$3.89''=468~AU), each candidate has a low likelihood of being bound, and so we did not collect follow-up data for this candidate. Candidates \#1, \#2 and \#3 are all fainter than 15th magnitude in H2 and have H2-H3 colors of 0.21, -0.11 and 0.34, and so we conclude all three are background objects. Candidate \#4 is too bright for H2-H3 color to differentiate between a background and a companion, but at a projected separation of 5.39''= 664~AU this object is highly likely to be a background star.

\hspace{-1em}\textit{HD~146606:} A single, faint candidate is identified at a separation of 4.7'', and at this wide separation is a likely background object. 

\hspace{-1em}\textit{HD~148657:} This target is just 6.8$^\circ$ from the galactic plane, and so there is a rich field of background objects. We identify a total of 29 candidates in our first epoch of data, 26 of which are redetected in a second epoch and confirmed to be background objects based on CPM and color analysis. The remaining three candidates are too faint to be identified in the second epoch. These three candidates are at relatively wide projected separations (161, 547 and 799~AU), and based on their faint H2 magnitudes and small H2-H3 colors, we conclude that all three are background objects.

\hspace{-1em}\textit{HD~151109:} We detect a total of 49 candidates around HD~151109, which is 4$^\circ$ from the galactic plane. 44 of these are redetected in a follow-up observation, and the remaining 5 are too faint to be detected in the second epoch. As discussed in Section \ref{multi-astrometry-text} above, the candidates we detect are systematically shifted by a smaller distance than would be expected based on the proper motion of this target. Based on the systematically similar motion of the set of candidates we conclude they are all likely background stars. The subset of candidates faint enough that H2-H3 color can be used to differentiate companions and background objects all have colors close to zero, confirming this assumption. The 5 candidates detected only in the first epoch are also highly likely to be background objects, based on their wide separation, faint absolute magnitude and small H2-H3 colors.

\hspace{-1em}\textit{HD~153053:} For this target, 14 candidates are identified and 13 of these are redetected in a second epoch of data and confirmed to be background objects. The final candidate, at a separation of 5.60'', is outside the field of view in the second epoch of data, due to the orientation of the ccamera. Based on the wide separation of this candidate, it is a likely background object.

\hspace{-1em}\textit{HD~182919:} A total of 40 candidates are detected around this target, which is 1.7$^\circ$ from the galactic plane. In a second epoch of data, we redetect 38 of these 40 candidates. As can be seen in Figure \ref{fig:gridfig_alls}, there is some scatter in final position relative to the predicted final positions for each candidate. We nonetheless conclude based on the proper motion, absolute magnitude, color and separation of each candidate that these are all background objects. The two candidates that appear only in the first epoch are also assumed to be background objects, based on their wide separation, faint absolute magnitude and small H2-H3 colors.

\hspace{-1em}\textit{HD~196544:} The two background objects identified in \citet{nielsen2013} are redetected, and no new candidates are found. 

\hspace{-1em}\textit{HD~215766:} No candidates are detected around this target.

\hspace{-1em}\textit{HD~223352:} This target was first identified as a tertiary system in \citet{derosa2011}, and redetected by \citet{rameau2013} and \citet{galicher2016}. We detect the companions HIP~117452Ba and HIP~117452Bb as listed in \citet{derosa2011}, but do not find any evidence for additional companions orbiting the primary, even with our improved contrast limits. Orbital motion of the binary pair relative to each other and to the primary is clearly detected relative to previous publications, and preliminary orbit fitting is now possible. This is beyond the scope of the current work, but we collate all published astrometry for the triple system in Table \ref{tab:223352}. \citet{zuckerman2011} list HD~223352 as a triple system, with a close binary and a tertiary object, HD~223340, an early-K-type at a separation of $\sim$75''. In this work and the other works referenced in Table \ref{tab:223352}, we resolve the binary of \citet{zuckerman2011} as \textit{three} distinct stars, meaning this system is in fact a \textit{quadruple} system with an A0 primary, orbited by a close binary pair at $\sim$3.5'' and additionally by a K-type star at $\sim$75''.

The binary pair is significantly more widely spaced than the debris gap: at $\sim$3.5'', it has a projected separation of 147~AU. We would therefore still expect a close-in planetary system between the two debris belts (at 4.6~AU and 41~AU) to be responsible for the dust clearing.
\label{sec:223352}

\hspace{-1em}\textit{HD~225200:} No candidates are detected around this target.

\begin{table*}
    \centering
    \caption{Astrometry for the two close companions of HD223352. The third companion, an early-K star at $\sim$75'', \citep{zuckerman2011} is outside the field of view.}
    \begin{tabular}{lllllc}

        Date & Sep ('') & $\upsigma_{\textrm{sep}}$ & PA & $\upsigma_{\textrm{PA}}$ & Reference \\
        \hline \hline
        \multicolumn{6}{c}{\textbf{HD223352Ba}} \\
        \hline
        2008-10-12 & 3.66   & 0.04   & 237.3  & --   & \citealt{galicher2016}  \\
        2009-08-30 & 3.7    & 0.1    & 237.3  & 0.4  & \citealt{derosa2011}   \\
        2009-12-31 & 3.67   & 0.04   & 237.3  & --   & \citealt{galicher2016}  \\
        2012-12-07 & 3.667  & 0.009  & 237.8  & 0.8  & \citealt{rameau2013}    \\
        2015-07-18 & 3.7141 & 0.0073 & 238.06 & 0.14 & This work             \\
        \hline
        \multicolumn{6}{c}{\textbf{HD223352Bb}} \\
        \hline
        2008-10-12 & 3.50   & 0.04   & 238.6  & --   & \citealt{galicher2016}  \\
        2009-08-30 & 3.5    & 0.1    & 238.5  & 0.5  & \citealt{derosa2011}   \\
        2009-12-31 & 3.48   & 0.04   & 239.0  & --   & \citealt{galicher2016}  \\
        2012-12-07 & 3.402  & 0.009  & 238.6  & 0.98 & \citealt{rameau2013}    \\
        2015-07-18 & 3.3738 & 0.0068 & 239.13 & 0.15 & This work             \\
        \hline
    \end{tabular}
    \label{tab:223352}
\end{table*}

\section{Analysis} \label{analysis}

The mass/radius parameter space for planets orbiting these 24 systems discussed in this work can now be tightly constrained by combining our VLT/SPHERE observations with dynamical arguments. Is is therefore possible to make inferences about the putative planetary systems hiding within the debris gaps.

\begin{figure*}
    \includegraphics[width=\textwidth,trim=1cm 1.5cm 2cm 0cm]{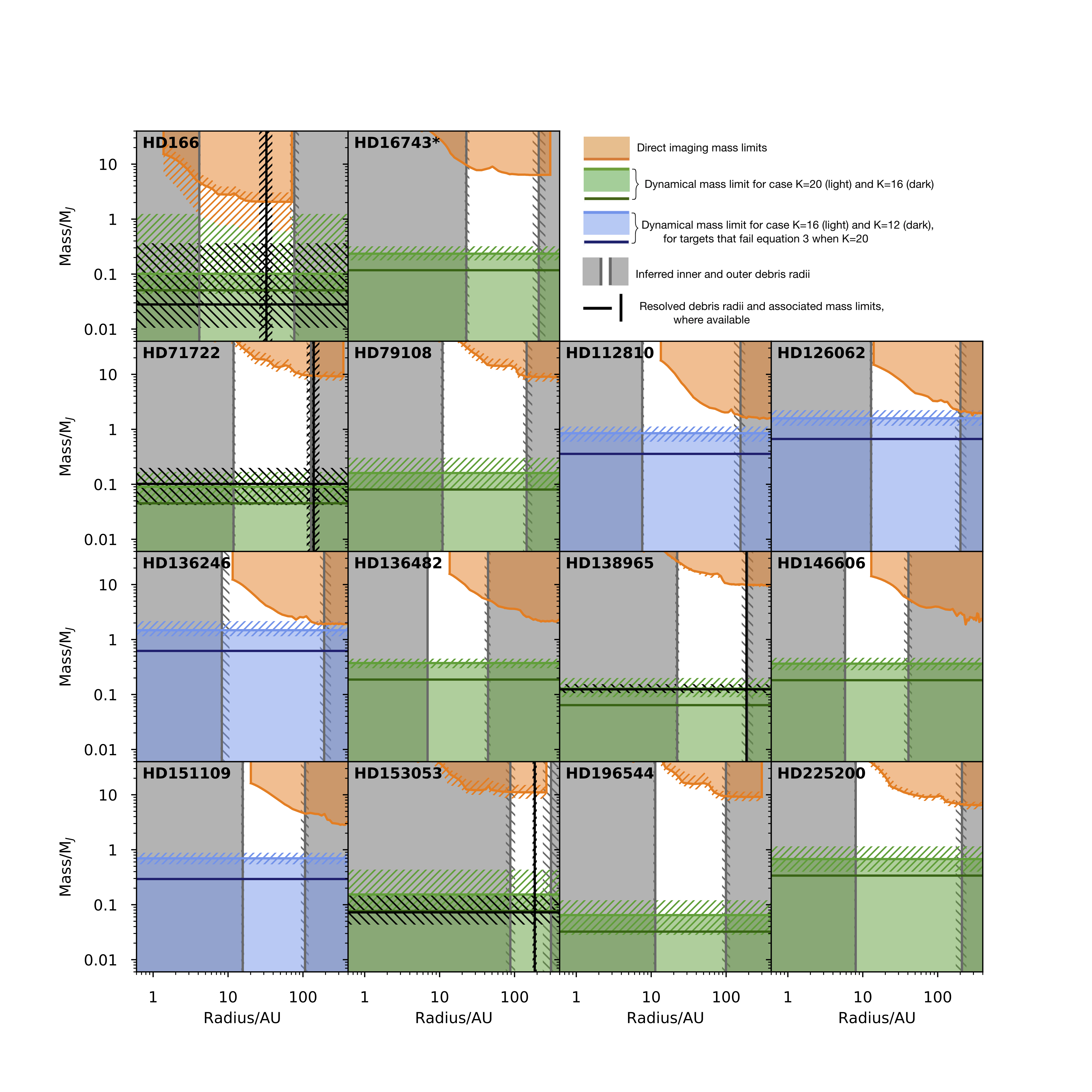} %,trim=1cm 1.5cm 2cm 0cm
    \caption{Constraints on the planetary systems for each of the targets in our survey. Since we do not detect companions, we expect the planetary systems to be within the white area of each subplot. The positions of the inner and outer debris belts are indicated in grey, with the regions inside the inner and beyond the outer shaded. Our direct imaging contrast limits based on SPHERE/IRDIS are shown in orange, with the region above this shaded, and dynamical mass constraints from \citet{shannon2016} are indicated in green, with masses below this value shaded. The uncertainty on this lower limit is calculated based on the age of host and the uncertainty in debris belt temperature, and indicated with hatching. Dark green lines indicate the lower limits for a slightly closer planet spacing of 16 mutual Hill radii. Errors are the same size as those on the light green lines but are not shown for clarity. For a small number of targets the inferred planetary mass from \citet{shannon2016} is too great to allow an interplanetary spacing of 20$R_H$. In these cases we instead show the 16$R_H$ case in blue, with an inferred 12$R_H$ limit shown in navy. In each case, a spacing of 12$R_H$ between each planet fulfils equation \ref{eq:massconstraint}. Additional black lines show the outer debris radius, and associated lower mass limit, for the subset of systems where the outer disk has been resolved (see Table \ref{tab:resolvedradii}). As mentioned Section \ref{sec:age}, for two of the targets (HD~16743 in this figure and HD~182919 in figure \ref{fig:massvshannon2}) the literature ages have no uncertainties, and we indicate these targets with asterisks.}
    \label{fig:massvshannon}
\end{figure*}

\begin{figure*}
    \includegraphics[width=\textwidth,trim=1cm 1.5cm 2cm 0cm]{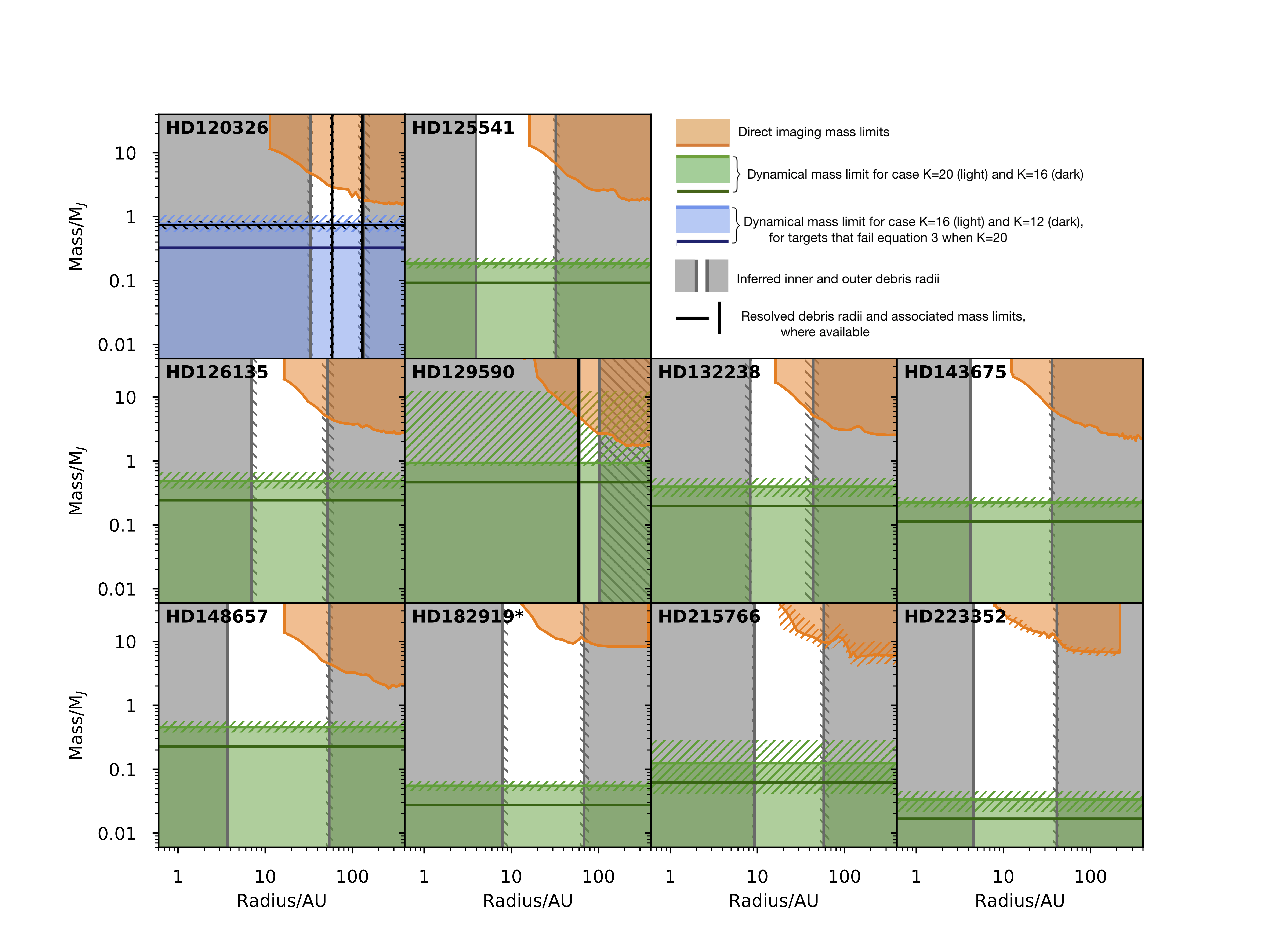}
    \caption{As for Figure \ref{fig:massvshannon}, but for candidates for which there is more uncertainty about the two-belt nature of the disks. In these cases we only present the two-belt planetary constraints.}
    \label{fig:massvshannon2}
\end{figure*}

Mass limits are calculated using the SPHERE/IRDIS contrast limits as described in Section \ref{sec:contrasts}: the COND models \citep{baraffe2003} are used for temperatures below 1700K and  the DUSTY models \citep{chabrier2000} otherwise. These mass limits are shown in Figures \ref{fig:massvshannon} and \ref{fig:massvshannon2}. For these mass limits, the confidence interval is calculated based solely on the age of the system. As discussed above in Section \ref{sec:age}, for two targets (HD~16743 and HD~182929) we were only able to find literature ages without uncertainties quoted, and as such are also unable to calculate uncertainties in our mass limits. For the 14 Sco-Cen targets in our sample, the ages are well determined \citep{pecaut2012} and so our uncertainties in mass limit are small. 

\label{sec:shannon-lower}
Also plotted in Figures \ref{fig:massvshannon} and \ref{fig:massvshannon2} are the minimum masses of planets required to clear the inferred debris gaps, based on the N-body simulations of \citet{shannon2016}. The quoted mass is the minimum mass \textit{per planet}, with uncertainties calculated based on the age of the system and the uncertainty in the disk radius. In all cases except for HD~129590, we infer that the system must be in a multi-planet configuration: the theoretical mass for a single planet to clear the gap is large ($\gtrsim$50$M_J$ in each case, \citealt{quillen2006,morrisonmalhotra2015}). \citet{quillen2006} found this clearing mass to be consistent for eccentricities $\lesssim$0.3. \citet{nesvoldkuchner2015} predicts slightly lower masses for a single gap-clearing planet, but still requires a single planet to have $\gtrsim$25$M_J$ to have cleared the observed gap. 

The minimum mass calculation relies on the assumption that planets are spaced by $\sim$20 mutual Hill radii. To demonstrate the impact of this choice of spacing, we additionally plot the lower mass limits for a spacing of 16$R_H$, as given in \citet{shannon2016}. The number of Hill radii between each planet has a relatively small effect on the dynamical constraints. For a planet spacing $>$20$R_H$, the predicted mass for each planet is higher than in the 20$R_H$ case, and so the lower limits that we plot remain valid.

In a small number of cases, the planet masses inferred from \citet{shannon2016} are sufficiently high that the value of $R_H$ approaches a significant fraction of the star-planet spacing. In this case, for planets to be spaced by $K R_H$, the second planet will be at semi-major axis $a_2 = a_1 + K R_H$. Substituting for equation \ref{eq:mutualhillradius}, the semi-major axis is given by
\begin{equation}
    a_2\left(1-\frac{K}{2}\left(\frac{m_1+m_2}{3M_\star}\right)^{\frac{1}{3}}\right) = a_1\left(1+\frac{K}{2}\left(\frac{m_1+m_2}{3M_\star}\right)^{\frac{1}{3}}\right),
\end{equation}
which only gives a positive value for the semi-major axis when
\begin{equation}
    m_1+m_2 < \frac{24M_\star}{K^3}.
    \label{eq:massconstraint}
\end{equation}
At this point, the very definition of mutual Hill radii means that for a specified planet mass, there is a certain maximum value of K, the number of $R_H$ between each planet. Alternatively, for a given K, there is a maximum mass of planets that fulfils equation \ref{eq:massconstraint}. For a solar mass star, and equal mass planets with K=20, this condition is reached at a planet mass of 1.43$M_J$. 

For a small number of the targets in this work, the masses inferred from \citet{shannon2016} are sufficiently high that this limit is reached, and equal-mass planets cannot be separated by 20 mutual Hill radii. For these targets, we instead calculate the lower limits for a spacing of 16$R_H$. These are highlighted in blue in Figure \ref{fig:massvshannon2}. When the number (K) of Hill radii between each target is changed, the clearing time appears to scale as K$^3$, so we tentatively also calculate a clearing time limit for a 12$R_H$ spacing as 0.42$\times$ the limit for 16$R_H$. This is an unusually close inter-planet spacing, and for all of the targets in our survey a value of K=12 predicts a clearing mass that satisfies equation \ref{eq:massconstraint}, and so a spacing of 12$R_H$ is reasonable.

We separate the targets for which there is doubt about the two-belt nature of the debris, and plot these in Figure \ref{fig:massvshannon2}. Our analysis is only valid if these are genuine two-belt systems.

By combining the observational upper and theoretical lower mass constraints in this way, only a small region of parameter space is left unconstrained. In some cases the region between the upper and lower mass constraints is less than an order of magnitude, with lower mass limits exceeding 1M$_J$ for systems with the widest debris disks and at the youngest ages. For all targets except HD~129590, we infer a multi-planet system based on the large theoretical clearing masses. In such a multi-planet system, the widest separation planet will have a physical separation close to that of the outer debris belt, where our direct imaging limits are relatively tight. Geometrical arguments mean that the planet will only appear at such a wide projected separation in a subset of cases, but this outermost planet is nonetheless constrained to a relatively small mass range, especially for targets where ALMA or \textit{Herschel} data constrains the system inclination.

In our survey of 24 targets, no exoplanetary mass companions were detected. For context, \citet{meshkat2017} found occurrence rates of 6.27\% (68\% confidence interval 3.68-9.76\%) in a debris disk sample of planets between 5-20$M_J$ and 10-1000~AU. Although our sample is too small for a detailed statistical analysis to be instructive, a non-detection in a sample of 24 stars is not inconsistent with the debris disk occurrence rate found in that work, since one would expect some companions might be geometrically unfavorably aligned, or below our detection limits. Our non-detections are also consistent with the lower occurence rate of $\sim$1\% found in unbiased samples by both \citet{bowler2016} and \citet{galicher2016}. The results of this survey are not incompatible with the theory that planets are carving wide debris gaps, since in each case our direct imaging upper mass limits are higher than the theoretical lower mass limits that we calculate. However, in several cases there is only a small mass range remaining where the planets could be massive enough to clear the observed gap, and yet sufficiently small and faint to remain undetected.

It is possible that the inferred gaps in these systems are not in fact caused by the presence of planets. An alternative cause of such a two-belt debris structure is that the belts form at the positions of molecular snow lines \citep[see e.g.][]{ballering2017,matra2018}, with the inner belt positioned at a water snow line and the outer belt at a CO snow line. The correlation found in \citet{kennedy2014} between the outer disk temperature and the stellar luminosity suggests that the dust location does not consistently match with a condensation temperature. However, in an optically thick disk the snow line positions would be determined by the mid plane temperature and so this correlation does not exclude the possibility of a more complex relationship between condensation positions and the formation of two-belt debris disks, and more work is needed to understand this possibility. It is also possible that the two temperatures in these debris disks do not correspond to two distinct radii of debris, as addressed in detail in \citet{kennedy2014}. However, the existence of the HR~8799 and HD~95086 systems where planets are known to reside in two-temperature debris disks and the solar system where planets are known to reside between two belts of debris implies that planets are a valid explanation for the formation of this debris structure in at least a subset of cases.

The very youngest systems are the most effective targets for a study like this one: in these cases the parameter space can be most tightly constrained. The ratio between the upper and lower limits for the  younger, more distant Sco-Cen stars in this survey is much smaller than that for the older, closer targets. For these younger, more distant targets, exoplanets are more luminous \citep[see e.g.][]{chabrier2000,baraffe2003}, although at these further distances the same absolute magnitude corresponds to a fainter apparent magnitude. Crucially, though, the lower limits inferred from \citet{shannon2016} are significantly higher in the case of younger targets, where gaps have only a limited time to form. This effect is so significant that even in the cases where the direct imaging mass limit is higher, constraints on the planetary system are still tighter for the youngest targets. 

\section{Conclusions} \label{conclusion}

In this work, we have imaged 24 debris disk hosting stars using the VLT/SPHERE instrument in IRDIFS mode. These targets were specifically selected as those that are likely to host multiple, segregated debris belts enclosing a debris gap. It is inferred that a system of one or more planets is responsible for the clearing of this wide debris gaps, as is the case for the solar system and for exoplanet hosts HR~8799 and HD~95086. We identify a total of 178 candidates. Two of these have been previously identified as companions, and the remainder are found to be background or likely background objects based on previous literature, common proper motion analysis, and the magnitude, color and separation of each candidate. Our survey reaches a typical contrast of $\sim$13mag at 0.25'' and $\sim$15mag at 1.0''. These contrasts are converted to mass limits for each target. We additionally calculate the \textit{minimum} required mass for planets in the system to have cleared the observed debris gap. Combining our upper and lower mass limits, we are able to tightly constrain the unexplored parameter space around these systems: typically, planets must be at least $\sim$0.2$M_J$ to clear the observed gap based on dynamical arguments, and in some cases the dynamical limit exceeds 1$M_J$. Direct imaging data from VLT/SPHERE, meanwhile, is sensitive to planets of $\sim$3.6$M_J$ for a typical target in our survey, and 1.7$M_J$ in the best case. Several of the inferred planetary systems will likely be detectable with the next generation of high contrast imagers.

\section*{Acknowledgements}

EM thanks the University of Exeter for support through a Ph.D. studentship. GMK is supported by the Royal Society as a Royal Society University Research Fellow. AS is partially supported by funding from the Center for Exoplanets and Habitable Worlds. The Center for Exoplanets and Habitable Worlds is supported by the Pennsylvania State University, the Eberly College of Science, and the Pennsylvania Space Grant Consortium.

This work is based on observations made with ESO Telescopes at the La Silla Paranal Observatory under programme IDs 095.C-0549, 095.C-0838, 097.C-0949, 097.C-1019, 099.C-0734. This work has made use of data from the European Space Agency (ESA) mission {\it Gaia} (\url{http://www.cosmos.esa.int/gaia}), processed by the {\it Gaia} Data Processing and Analysis Consortium (DPAC, \url{http://www.cosmos.esa.int/web/gaia/dpac/consortium}). Funding for the DPAC has been provided by national institutions, in particular the institutions participating in the {\it Gaia} Multilateral Agreement. This research has made use of the SIMBAD database, operated at CDS, Strasbourg, France.

\bibliographystyle{mnras}
\bibliography{bibliography.bib}

\begin{thebibliography}{}
\makeatletter
\relax
\def\mn@urlcharsother{\let\do\@makeother \do\$\do\&\do\#\do\^\do\_\do\%\do\~}
\def\mn@doi{\begingroup\mn@urlcharsother \@ifnextchar [ {\mn@doi@}
  {\mn@doi@[]}}
\def\mn@doi@[#1]#2{\def\@tempa{#1}\ifx\@tempa\@empty \href
  {http://dx.doi.org/#2} {doi:#2}\else \href {http://dx.doi.org/#2} {#1}\fi
  \endgroup}
\def\mn@eprint#1#2{\mn@eprint@#1:#2::\@nil}
\def\mn@eprint@arXiv#1{\href {http://arxiv.org/abs/#1} {{\tt arXiv:#1}}}
\def\mn@eprint@dblp#1{\href {http://dblp.uni-trier.de/rec/bibtex/#1.xml}
  {dblp:#1}}
\def\mn@eprint@#1:#2:#3:#4\@nil{\def\@tempa {#1}\def\@tempb {#2}\def\@tempc
  {#3}\ifx \@tempc \@empty \let \@tempc \@tempb \let \@tempb \@tempa \fi \ifx
  \@tempb \@empty \def\@tempb {arXiv}\fi \@ifundefined
  {mn@eprint@\@tempb}{\@tempb:\@tempc}{\expandafter \expandafter \csname
  mn@eprint@\@tempb\endcsname \expandafter{\@tempc}}}

\bibitem[\protect\citeauthoryear{{Amara} \& {Quanz}}{{Amara} \&
  {Quanz}}{2012}]{amaraquanz2012}
{Amara} A.,  {Quanz} S.~P.,  2012, \mnras, 427, 948

\bibitem[\protect\citeauthoryear{{Ballering}, {Rieke}, {Su}  \&
  {Montiel}}{{Ballering} et~al.}{2013}]{ballering2013}
{Ballering} N.~P.,  {Rieke} G.~H.,  {Su} K.~Y.~L.,   {Montiel} E.,  2013, \apj,
  775, 55

\bibitem[\protect\citeauthoryear{{Ballering}, {Rieke}, {Su}  \&
  {G{\'a}sp{\'a}r}}{{Ballering} et~al.}{2017}]{ballering2017}
{Ballering} N.~P.,  {Rieke} G.~H.,  {Su} K.~Y.~L.,   {G{\'a}sp{\'a}r} A.,
  2017, \mn@doi [\apj] {10.3847/1538-4357/aa8037}, \href
  {http://adsabs.harvard.edu/abs/2017ApJ...845..120B} {845, 120}

\bibitem[\protect\citeauthoryear{{Baraffe}, {Chabrier}, {Barman}, {Allard}  \&
  {Hauschildt}}{{Baraffe} et~al.}{2003}]{baraffe2003}
{Baraffe} I.,  {Chabrier} G.,  {Barman} T.~S.,  {Allard} F.,   {Hauschildt}
  P.~H.,  2003, \aap, 402, 701

\bibitem[\protect\citeauthoryear{Beuzit et~al.,}{Beuzit
  et~al.}{2008}]{beuzit2008}
Beuzit J.-L.,  et~al., 2008, \spie, 7014, 701418

\bibitem[\protect\citeauthoryear{{Biller} et~al.,}{{Biller}
  et~al.}{2013}]{biller2013}
{Biller} B.~A.,  et~al., 2013, \apj, 777, 160

\bibitem[\protect\citeauthoryear{{Bonnefoy} et~al.,}{{Bonnefoy}
  et~al.}{2017}]{bonnefoy2017}
{Bonnefoy} M.,  et~al., 2017, \aap, 597, L7

\bibitem[\protect\citeauthoryear{{Bowler}}{{Bowler}}{2016}]{bowler2016}
{Bowler} B.~P.,  2016, \pasp, 128, 102001

\bibitem[\protect\citeauthoryear{{Brandt} \& {Huang}}{{Brandt} \&
  {Huang}}{2015}]{brandt2015}
{Brandt} T.~D.,  {Huang} C.~X.,  2015, \apj, 807, 58

\bibitem[\protect\citeauthoryear{{Brandt} et~al.,}{{Brandt}
  et~al.}{2014}]{brandt2014}
{Brandt} T.~D.,  et~al., 2014, \apj, 794, 159

\bibitem[\protect\citeauthoryear{{Chabrier}, {Baraffe}, {Allard}  \&
  {Hauschildt}}{{Chabrier} et~al.}{2000}]{chabrier2000}
{Chabrier} G.,  {Baraffe} I.,  {Allard} F.,   {Hauschildt} P.,  2000, \apj,
  542, 464

\bibitem[\protect\citeauthoryear{{Chauvin} et~al.,}{{Chauvin}
  et~al.}{2015}]{chauvin2015}
{Chauvin} G.,  et~al., 2015, \aap, 573, A127

\bibitem[\protect\citeauthoryear{{Chauvin} et~al.,}{{Chauvin}
  et~al.}{2017}]{chauvin2017}
{Chauvin} G.,  et~al., 2017, \aap, 605, L9

\bibitem[\protect\citeauthoryear{{Chen}, {Mittal}, {Kuchner}, {Forrest},
  {Lisse}, {Manoj}, {Sargent}  \& {Watson}}{{Chen} et~al.}{2014}]{chen2014}
{Chen} C.~H.,  {Mittal} T.,  {Kuchner} M.,  {Forrest} W.~J.,  {Lisse} C.~M.,
  {Manoj} P.,  {Sargent} B.~A.,   {Watson} D.~M.,  2014, \apjs, 211, 25

\bibitem[\protect\citeauthoryear{Claudi et~al.,}{Claudi
  et~al.}{2008}]{claudi2008}
Claudi R.~U.,  et~al., 2008, \spie, 7014, 70143

\bibitem[\protect\citeauthoryear{{Currie}, {Lisse}, {Kuchner}, {Madhusudhan},
  {Kenyon}, {Thalmann}, {Carson}  \& {Debes}}{{Currie}
  et~al.}{2015}]{currie2015b}
{Currie} T.,  {Lisse} C.~M.,  {Kuchner} M.,  {Madhusudhan} N.,  {Kenyon} S.~J.,
   {Thalmann} C.,  {Carson} J.,   {Debes} J.,  2015, \apjl, 807, L7

\bibitem[\protect\citeauthoryear{{David} \& {Hillenbrand}}{{David} \&
  {Hillenbrand}}{2015}]{davidhillenbrand2015}
{David} T.~J.,  {Hillenbrand} L.~A.,  2015, \apj, 804, 146

\bibitem[\protect\citeauthoryear{{De Rosa} et~al.,}{{De Rosa}
  et~al.}{2011}]{derosa2011}
{De Rosa} R.~J.,  et~al., 2011, \mnras, 415, 854

\bibitem[\protect\citeauthoryear{{De Rosa} et~al.,}{{De Rosa}
  et~al.}{2015}]{derosa2015}
{De Rosa} R.~J.,  et~al., 2015, \apjl, 814, L3

\bibitem[\protect\citeauthoryear{Dohlen et~al.,}{Dohlen
  et~al.}{2008}]{dohlen2008}
Dohlen K.,  et~al., 2008, \spie, 7014, 70143

\bibitem[\protect\citeauthoryear{{Draper} et~al.,}{{Draper}
  et~al.}{2016}]{draper2016}
{Draper} Z.~H.,  et~al., 2016, \apj, 826, 147

\bibitem[\protect\citeauthoryear{{Fabrycky} \& {Murray-Clay}}{{Fabrycky} \&
  {Murray-Clay}}{2010}]{fabryckymurrayclay2010}
{Fabrycky} D.~C.,  {Murray-Clay} R.~A.,  2010, \apj, 710, 1408

\bibitem[\protect\citeauthoryear{{Fang} \& {Margot}}{{Fang} \&
  {Margot}}{2013}]{fangmargot2013}
{Fang} J.,  {Margot} J.-L.,  2013, \apj, 767, 115

\bibitem[\protect\citeauthoryear{{Feldt} et~al.,}{{Feldt}
  et~al.}{2017}]{feldt2017}
{Feldt} M.,  et~al., 2017, \aap, 601, A7

\bibitem[\protect\citeauthoryear{{Gaia Collaboration} et~al.,}{{Gaia
  Collaboration} et~al.}{2016a}]{gaia_dr1b}
{Gaia Collaboration} et~al., 2016a, \aap, 595, A1

\bibitem[\protect\citeauthoryear{{Gaia Collaboration} et~al.,}{{Gaia
  Collaboration} et~al.}{2016b}]{gaia_dr1a}
{Gaia Collaboration} et~al., 2016b, \aap, 595, A2

\bibitem[\protect\citeauthoryear{{Galicher} et~al.,}{{Galicher}
  et~al.}{2016}]{galicher2016}
{Galicher} R.,  et~al., 2016, \aap, 594, A63

\bibitem[\protect\citeauthoryear{{Gerbaldi}, {Faraggiana}, {Burnage}, {Delmas},
  {G{\'o}mez}  \& {Grenier}}{{Gerbaldi} et~al.}{1999}]{gerbaldi1999}
{Gerbaldi} M.,  {Faraggiana} R.,  {Burnage} R.,  {Delmas} F.,  {G{\'o}mez}
  A.~E.,   {Grenier} S.,  1999, \aaps, 137, 273

\bibitem[\protect\citeauthoryear{{Go{\'z}dziewski} \&
  {Migaszewski}}{{Go{\'z}dziewski} \&
  {Migaszewski}}{2014}]{gozdziewskimigaszewski2014}
{Go{\'z}dziewski} K.,  {Migaszewski} C.,  2014, \mnras, 440, 3140

\bibitem[\protect\citeauthoryear{{Janson} et~al.,}{{Janson}
  et~al.}{2013}]{janson2013a}
{Janson} M.,  et~al., 2013, \apj, 773, 73

\bibitem[\protect\citeauthoryear{{Kasper}, {Apai}, {Wagner}  \&
  {Robberto}}{{Kasper} et~al.}{2015}]{kasper2015}
{Kasper} M.,  {Apai} D.,  {Wagner} K.,   {Robberto} M.,  2015, \apjl, 812, L33

\bibitem[\protect\citeauthoryear{{Kennedy} \& {Wyatt}}{{Kennedy} \&
  {Wyatt}}{2014}]{kennedy2014}
{Kennedy} G.~M.,  {Wyatt} M.~C.,  2014, \mnras, 444, 3164

\bibitem[\protect\citeauthoryear{{Keppler} et~al.,}{{Keppler}
  et~al.}{2018}]{keppler2018}
{Keppler} M.,  et~al., 2018, preprint, \href
  {http://adsabs.harvard.edu/abs/2018arXiv180611568K} {} (\mn@eprint {arXiv}
  {1806.11568})

\bibitem[\protect\citeauthoryear{{Lafreni{\`e}re} et~al.,}{{Lafreni{\`e}re}
  et~al.}{2007}]{lafreniere2007}
{Lafreni{\`e}re} D.,  et~al., 2007, \apj, 670, 1367

\bibitem[\protect\citeauthoryear{{Lagrange} et~al.,}{{Lagrange}
  et~al.}{2009}]{lagrange2009}
{Lagrange} A.-M.,  et~al., 2009, \aap, 493, L21

\bibitem[\protect\citeauthoryear{{Leconte} et~al.,}{{Leconte}
  et~al.}{2010}]{leconte2010}
{Leconte} J.,  et~al., 2010, \apj, 716, 1551

\bibitem[\protect\citeauthoryear{{L{\'o}pez-Santiago}, {Montes},
  {Crespo-Chac{\'o}n}  \& {Fern{\'a}ndez-Figueroa}}{{L{\'o}pez-Santiago}
  et~al.}{2006}]{lopezsantiago2006}
{L{\'o}pez-Santiago} J.,  {Montes} D.,  {Crespo-Chac{\'o}n} I.,
  {Fern{\'a}ndez-Figueroa} M.~J.,  2006, \apj, 643, 1160

\bibitem[\protect\citeauthoryear{{Lynga} \& {Wramdemark}}{{Lynga} \&
  {Wramdemark}}{1984}]{lynga1984}
{Lynga} G.,  {Wramdemark} S.,  1984, \aap, 132, 58

\bibitem[\protect\citeauthoryear{{Macintosh} et~al.,}{{Macintosh}
  et~al.}{2014}]{macintosh2014}
{Macintosh} B.,  et~al., 2014, \pnas, 111, 12661

\bibitem[\protect\citeauthoryear{{Macintosh} et~al.,}{{Macintosh}
  et~al.}{2015}]{macintosh2015}
{Macintosh} B.,  et~al., 2015, \sci, 350, 64

\bibitem[\protect\citeauthoryear{{Maire} et~al.,}{{Maire}
  et~al.}{2016}]{maire2016}
{Maire} A.-L.,  et~al., 2016, \spie, 9908, 990834

\bibitem[\protect\citeauthoryear{{Maldonado}, {Mart{\'{\i}}nez-Arn{\'a}iz},
  {Eiroa}, {Montes}  \& {Montesinos}}{{Maldonado} et~al.}{2010}]{maldonado2010}
{Maldonado} J.,  {Mart{\'{\i}}nez-Arn{\'a}iz} R.~M.,  {Eiroa} C.,  {Montes} D.,
    {Montesinos} B.,  2010, \aap, 521, A12

\bibitem[\protect\citeauthoryear{{Mamajek}}{{Mamajek}}{2016}]{mamajek2016}
{Mamajek} E.~E.,  2016, IAU Proc., 314, 21

\bibitem[\protect\citeauthoryear{Marois, Lafreni{\`e}re, Doyon, Macintosh  \&
  Nadeau}{Marois et~al.}{2006}]{marois2006}
Marois C.,  Lafreni{\`e}re D.,  Doyon R.,  Macintosh B.,   Nadeau D.,  2006,
  \apj, 641, 556

\bibitem[\protect\citeauthoryear{{Marois}, {Macintosh}, {Barman}, {Zuckerman},
  {Song}, {Patience}, {Lafreni{\`e}re}  \& {Doyon}}{{Marois}
  et~al.}{2008}]{marois2008}
{Marois} C.,  {Macintosh} B.,  {Barman} T.,  {Zuckerman} B.,  {Song} I.,
  {Patience} J.,  {Lafreni{\`e}re} D.,   {Doyon} R.,  2008, \sci, 322, 1348

\bibitem[\protect\citeauthoryear{{Marois}, {Zuckerman}, {Konopacky},
  {Macintosh}  \& {Barman}}{{Marois} et~al.}{2010}]{marois2010}
{Marois} C.,  {Zuckerman} B.,  {Konopacky} Q.~M.,  {Macintosh} B.,   {Barman}
  T.,  2010, \nat, 468, 1080

\bibitem[\protect\citeauthoryear{{Matr{\`a}}, {Marino}, {Kennedy}, {Wyatt},
  {{\"O}berg}  \& {Wilner}}{{Matr{\`a}} et~al.}{2018}]{matra2018}
{Matr{\`a}} L.,  {Marino} S.,  {Kennedy} G.~M.,  {Wyatt} M.~C.,  {{\"O}berg}
  K.~I.,   {Wilner} D.~J.,  2018, \mn@doi [\apj] {10.3847/1538-4357/aabcc4},
  \href {http://adsabs.harvard.edu/abs/2018ApJ...859...72M} {859, 72}

\bibitem[\protect\citeauthoryear{{Matthews}, {Kennedy}, {Sibthorpe}, {Booth},
  {Wyatt}, {Broekhoven-Fiene}, {Macintosh}  \& {Marois}}{{Matthews}
  et~al.}{2014}]{matthews_b2014}
{Matthews} B.,  {Kennedy} G.,  {Sibthorpe} B.,  {Booth} M.,  {Wyatt} M.,
  {Broekhoven-Fiene} H.,  {Macintosh} B.,   {Marois} C.,  2014, \apj, 780, 97

\bibitem[\protect\citeauthoryear{{Matthews} et~al.,}{{Matthews}
  et~al.}{2017}]{matthews2017}
{Matthews} E.,  et~al., 2017, \apjl, 843, L12

\bibitem[\protect\citeauthoryear{{Mawet} et~al.,}{{Mawet}
  et~al.}{2014}]{mawet2014}
{Mawet} D.,  et~al., 2014, \apj, 792, 97

\bibitem[\protect\citeauthoryear{{Mesa} et~al.,}{{Mesa}
  et~al.}{2015}]{mesa2016}
{Mesa} D.,  et~al., 2015, \aap, 576, A121

\bibitem[\protect\citeauthoryear{{Meshkat} et~al.,}{{Meshkat}
  et~al.}{2017}]{meshkat2017}
{Meshkat} T.,  et~al., 2017, \aj, 154, 245

\bibitem[\protect\citeauthoryear{{Morales}, {Rieke}, {Werner}, {Bryden},
  {Stapelfeldt}  \& {Su}}{{Morales} et~al.}{2011}]{morales2011}
{Morales} F.~Y.,  {Rieke} G.~H.,  {Werner} M.~W.,  {Bryden} G.,  {Stapelfeldt}
  K.~R.,   {Su} K.~Y.~L.,  2011, \apjl, 730, L29

\bibitem[\protect\citeauthoryear{{Morales}, {Bryden}, {Werner}  \&
  {Stapelfeldt}}{{Morales} et~al.}{2016}]{morales2016}
{Morales} F.~Y.,  {Bryden} G.,  {Werner} M.~W.,   {Stapelfeldt} K.~R.,  2016,
  \apj, 831, 97

\bibitem[\protect\citeauthoryear{{Morrison} \& {Malhotra}}{{Morrison} \&
  {Malhotra}}{2015}]{morrisonmalhotra2015}
{Morrison} S.,  {Malhotra} R.,  2015, \apj, 799, 41

\bibitem[\protect\citeauthoryear{{Mustill} \& {Wyatt}}{{Mustill} \&
  {Wyatt}}{2009}]{mustillwyatt2009}
{Mustill} A.~J.,  {Wyatt} M.~C.,  2009, \mnras, 399, 1403

\bibitem[\protect\citeauthoryear{{Nakajima} \& {Morino}}{{Nakajima} \&
  {Morino}}{2012}]{nakajima-morino2012}
{Nakajima} T.,  {Morino} J.-I.,  2012, \aj, 143, 2

\bibitem[\protect\citeauthoryear{{Nesvold} \& {Kuchner}}{{Nesvold} \&
  {Kuchner}}{2015}]{nesvoldkuchner2015}
{Nesvold} E.~R.,  {Kuchner} M.~J.,  2015, \apj, 798, 83

\bibitem[\protect\citeauthoryear{{Nielsen} et~al.,}{{Nielsen}
  et~al.}{2013}]{nielsen2013}
{Nielsen} E.~L.,  et~al., 2013, \apj, 776, 4

\bibitem[\protect\citeauthoryear{{Padgett} \& {Stapelfeldt}}{{Padgett} \&
  {Stapelfeldt}}{2016}]{padgett2016}
{Padgett} D.,  {Stapelfeldt} K.,  2016, IAU Proc., 314, 175

\bibitem[\protect\citeauthoryear{{Panagi} \& {O'dell}}{{Panagi} \&
  {O'dell}}{1997}]{panagi1997}
{Panagi} P.~M.,  {O'dell} M.~A.,  1997, \aaps, 121

\bibitem[\protect\citeauthoryear{{Pavlov}, {M{\"o}ller-Nilsson}, {Feldt},
  {Henning}, {Beuzit}  \& {Mouillet}}{{Pavlov} et~al.}{2008}]{pavlov2008}
{Pavlov} A.,  {M{\"o}ller-Nilsson} O.,  {Feldt} M.,  {Henning} T.,  {Beuzit}
  J.-L.,   {Mouillet} D.,  2008, \spie, 7019, 701939

\bibitem[\protect\citeauthoryear{{Pawellek} \& {Krivov}}{{Pawellek} \&
  {Krivov}}{2015}]{pawellek2015}
{Pawellek} N.,  {Krivov} A.~V.,  2015, \mnras, 454, 3207

\bibitem[\protect\citeauthoryear{{Pecaut} \& {Mamajek}}{{Pecaut} \&
  {Mamajek}}{2016}]{pecautmamajek2016}
{Pecaut} M.~J.,  {Mamajek} E.~E.,  2016, \mnras, 461, 794

\bibitem[\protect\citeauthoryear{{Pecaut}, {Mamajek}  \& {Bubar}}{{Pecaut}
  et~al.}{2012}]{pecaut2012}
{Pecaut} M.~J.,  {Mamajek} E.~E.,   {Bubar} E.~J.,  2012, \apj, 746, 154

\bibitem[\protect\citeauthoryear{{Perryman} et~al.,}{{Perryman}
  et~al.}{1997}]{hipparcos}
{Perryman} M.~A.~C.,  et~al., 1997, \aap, 323

\bibitem[\protect\citeauthoryear{{Quillen}}{{Quillen}}{2006}]{quillen2006}
{Quillen} A.~C.,  2006, \mnras, 372, L14

\bibitem[\protect\citeauthoryear{{Rameau} et~al.,}{{Rameau}
  et~al.}{2013}]{rameau2013}
{Rameau} J.,  et~al., 2013, \apjl, 779, L26

\bibitem[\protect\citeauthoryear{{Reidemeister}, {Krivov}, {Schmidt},
  {Fiedler}, {M{\"u}ller}, {L{\"o}hne}  \& {Neuh{\"a}user}}{{Reidemeister}
  et~al.}{2009}]{reidemeister2009}
{Reidemeister} M.,  {Krivov} A.~V.,  {Schmidt} T.~O.~B.,  {Fiedler} S.,
  {M{\"u}ller} S.,  {L{\"o}hne} T.,   {Neuh{\"a}user} R.,  2009, \aap, 503, 247

\bibitem[\protect\citeauthoryear{{Rhee}, {Song}, {Zuckerman}  \&
  {McElwain}}{{Rhee} et~al.}{2007}]{rhee07}
{Rhee} J.~H.,  {Song} I.,  {Zuckerman} B.,   {McElwain} M.,  2007, \apj, 660,
  1556

\bibitem[\protect\citeauthoryear{{Shannon}, {Bonsor}, {Kral}  \&
  {Matthews}}{{Shannon} et~al.}{2016}]{shannon2016}
{Shannon} A.,  {Bonsor} A.,  {Kral} Q.,   {Matthews} E.,  2016, \mnras, 462,
  L116

\bibitem[\protect\citeauthoryear{{Smith} \& {Terrile}}{{Smith} \&
  {Terrile}}{1984}]{smith1984}
{Smith} B.~A.,  {Terrile} R.~J.,  1984, \sci, 226, 1421

\bibitem[\protect\citeauthoryear{{Soummer}, {Pueyo}  \& {Larkin}}{{Soummer}
  et~al.}{2012}]{soummer2012}
{Soummer} R.,  {Pueyo} L.,   {Larkin} J.,  2012, \apjl, 755, L28

\bibitem[\protect\citeauthoryear{{Su} et~al.,}{{Su} et~al.}{2009}]{su2009}
{Su} K.~Y.~L.,  et~al., 2009, \apj, 705, 314

\bibitem[\protect\citeauthoryear{{Su}, {Morrison}, {Malhotra}, {Smith}, {Balog}
   \& {Rieke}}{{Su} et~al.}{2015}]{su2015}
{Su} K.~Y.~L.,  {Morrison} S.,  {Malhotra} R.,  {Smith} P.~S.,  {Balog} Z.,
  {Rieke} G.~H.,  2015, \apj, 799, 146

\bibitem[\protect\citeauthoryear{{Tetzlaff}, {Neuh{\"a}user}  \&
  {Hohle}}{{Tetzlaff} et~al.}{2011}]{tetzlaff2011}
{Tetzlaff} N.,  {Neuh{\"a}user} R.,   {Hohle} M.~M.,  2011, \mnras, 410, 190

\bibitem[\protect\citeauthoryear{{Vigan}, {Moutou}, {Langlois}, {Allard},
  {Boccaletti}, {Carbillet}, {Mouillet}  \& {Smith}}{{Vigan}
  et~al.}{2010}]{vigan2010}
{Vigan} A.,  {Moutou} C.,  {Langlois} M.,  {Allard} F.,  {Boccaletti} A.,
  {Carbillet} M.,  {Mouillet} D.,   {Smith} I.,  2010, \mnras, 407, 71

\bibitem[\protect\citeauthoryear{{Vigan}, {Gry}, {Salter}, {Mesa}, {Homeier},
  {Moutou}  \& {Allard}}{{Vigan} et~al.}{2015}]{vigan2015sirius}
{Vigan} A.,  {Gry} C.,  {Salter} G.,  {Mesa} D.,  {Homeier} D.,  {Moutou} C.,
  {Allard} F.,  2015, \mnras, 454, 129

\bibitem[\protect\citeauthoryear{{Wahhaj} et~al.,}{{Wahhaj}
  et~al.}{2013}]{wahhaj2013}
{Wahhaj} Z.,  et~al., 2013, \apj, 773, 179

\bibitem[\protect\citeauthoryear{{Wahhaj} et~al.,}{{Wahhaj}
  et~al.}{2016}]{wahhaj2016}
{Wahhaj} Z.,  et~al., 2016, \aap, 596, L4

\bibitem[\protect\citeauthoryear{{Zorec} \& {Royer}}{{Zorec} \&
  {Royer}}{2012a}]{zorecroyer2012}
{Zorec} J.,  {Royer} F.,  2012a, \aap, 537, A120

\bibitem[\protect\citeauthoryear{{Zorec} \& {Royer}}{{Zorec} \&
  {Royer}}{2012b}]{zorec2012}
{Zorec} J.,  {Royer} F.,  2012b, \aap, 537, A120

\bibitem[\protect\citeauthoryear{{Zuckerman}, {Rhee}, {Song}  \&
  {Bessell}}{{Zuckerman} et~al.}{2011}]{zuckerman2011}
{Zuckerman} B.,  {Rhee} J.~H.,  {Song} I.,   {Bessell} M.~S.,  2011, \apj, 732,
  61

\bibitem[\protect\citeauthoryear{{Zurlo} et~al.,}{{Zurlo}
  et~al.}{2014}]{zurlo2014}
{Zurlo} A.,  et~al., 2014, \aap, 572, A85

\bibitem[\protect\citeauthoryear{{de Zeeuw}, {Hoogerwerf}, {de Bruijne},
  {Brown}  \& {Blaauw}}{{de Zeeuw} et~al.}{1999}]{dezeeuw1999}
{de Zeeuw} P.~T.,  {Hoogerwerf} R.,  {de Bruijne} J.~H.~J.,  {Brown} A.~G.~A.,
   {Blaauw} A.,  1999, \aj, 117, 354

\makeatother
\end{thebibliography}

\onecolumn
\appendix

\section{Common Proper Motion plots}
\label{app:cpm}

Here we present the Common Proper Motion plots for each candidate with multiple epochs of data. In each case, we present our astrometry of the candidate relative to the host, along with any previous astrometry of the candidate from the literature. Dark points indicate the measured position, and light points indicate the predicted position for a background object at each epoch. A black line traces the path that a stationary background object would trace out relative to the host. This plot is an expansion of Figure \ref{fig:gridfig_comparative} in the main text.

\begin{figure}
	\includegraphics[width=\textwidth]{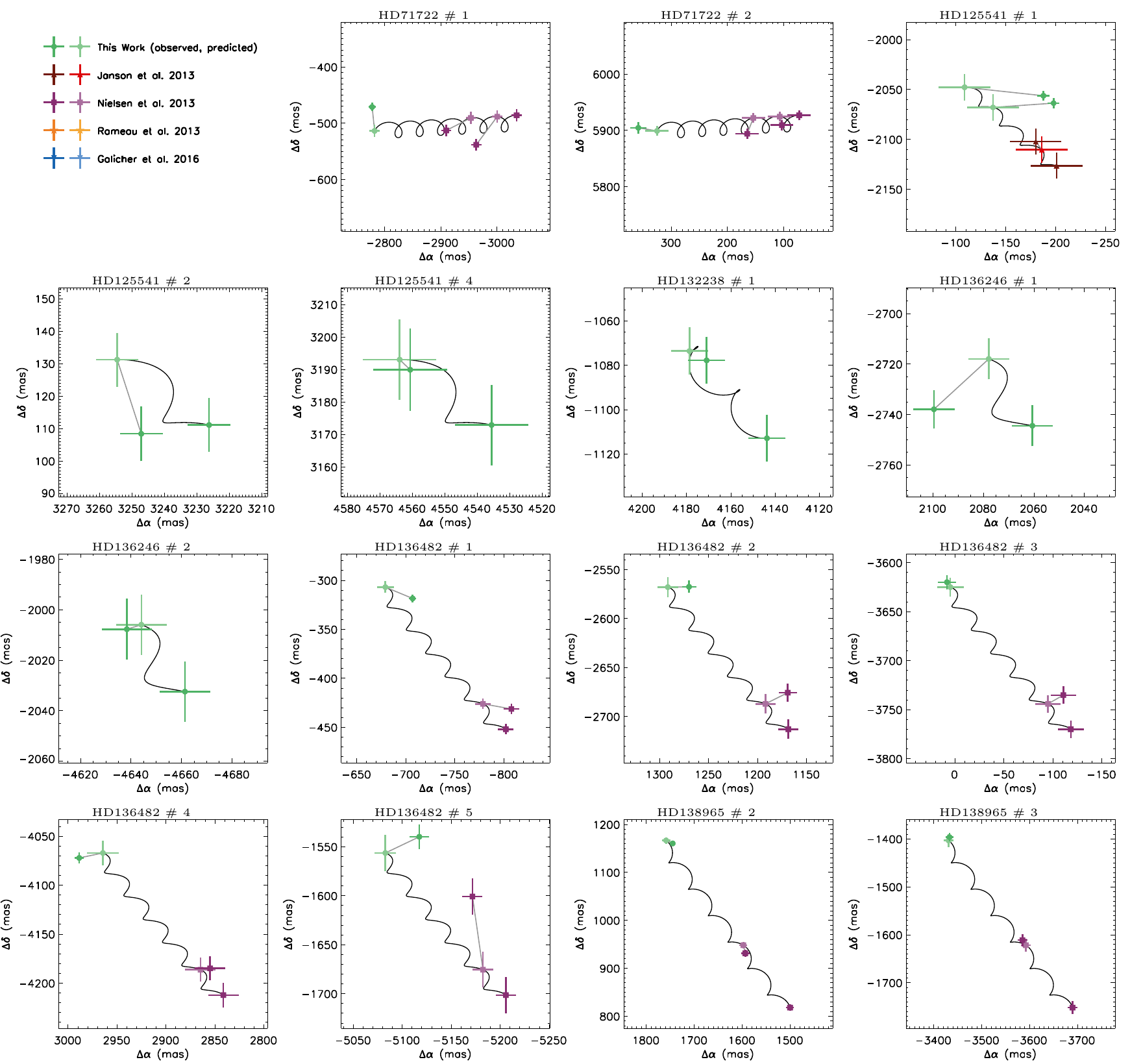}
\end{figure}

\begin{figure*}
	\includegraphics[width=\textwidth]{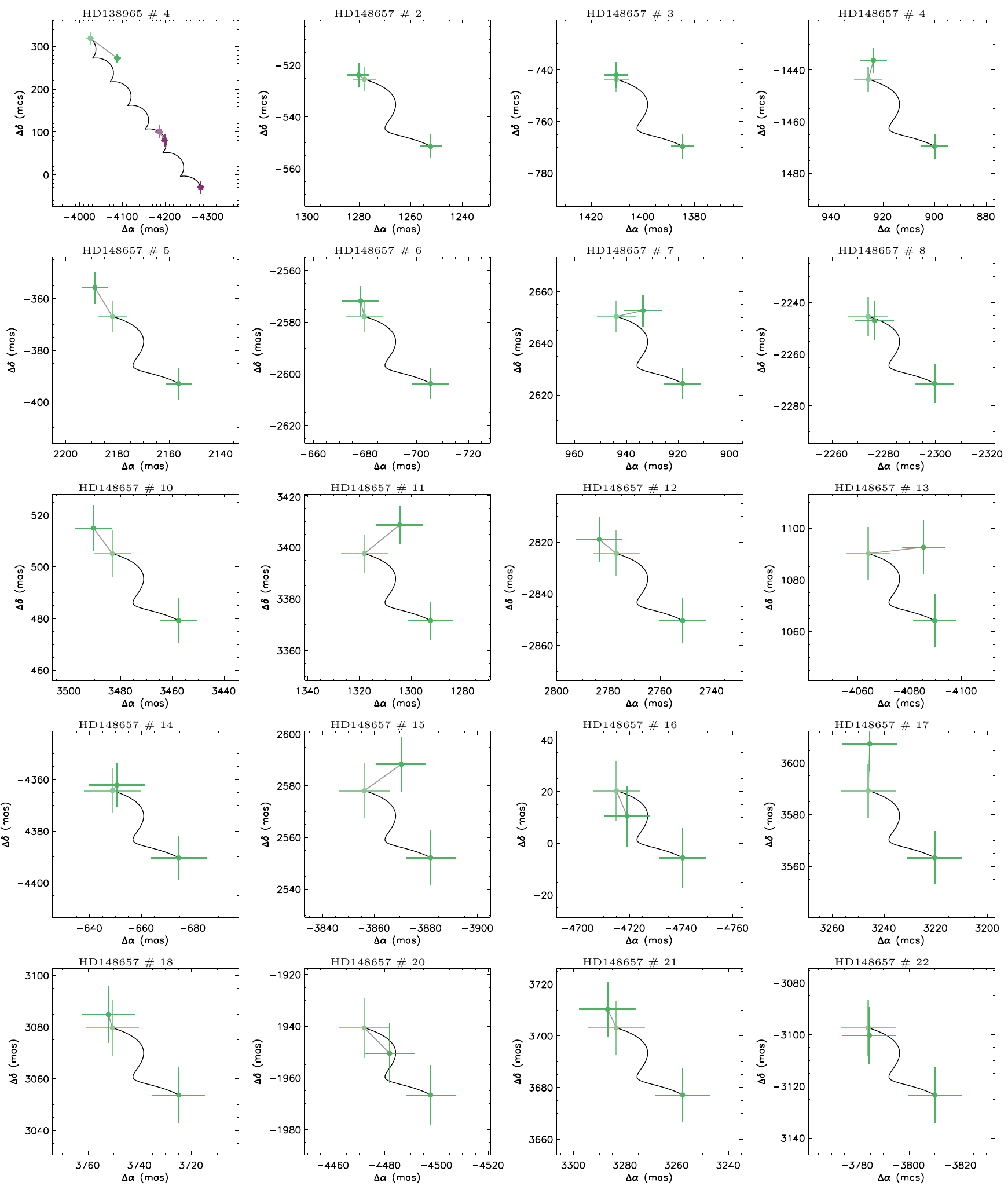}
\end{figure*}

\begin{figure*}
	\includegraphics[width=\textwidth]{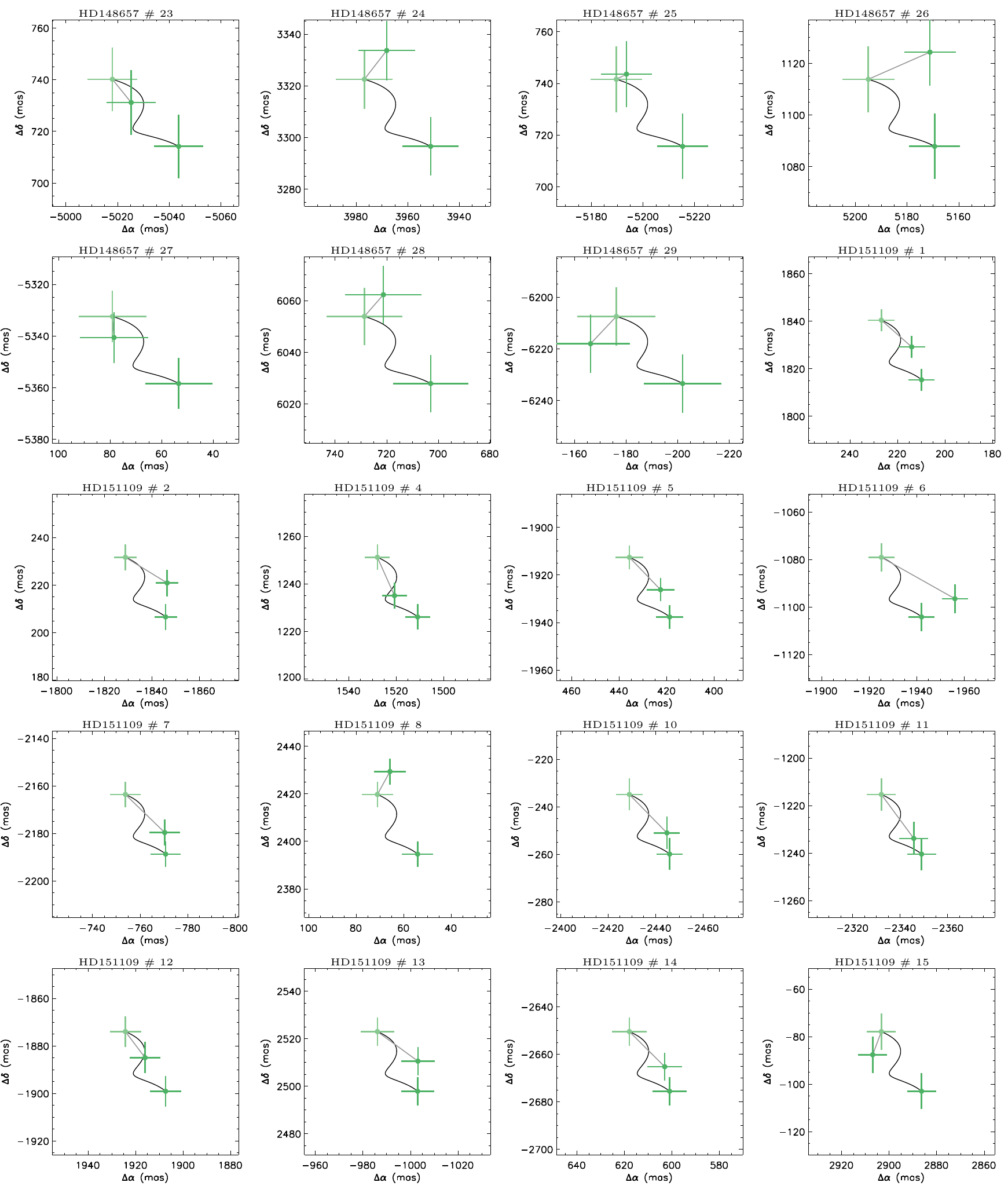}
\end{figure*}

\begin{figure*}
	\includegraphics[width=\textwidth]{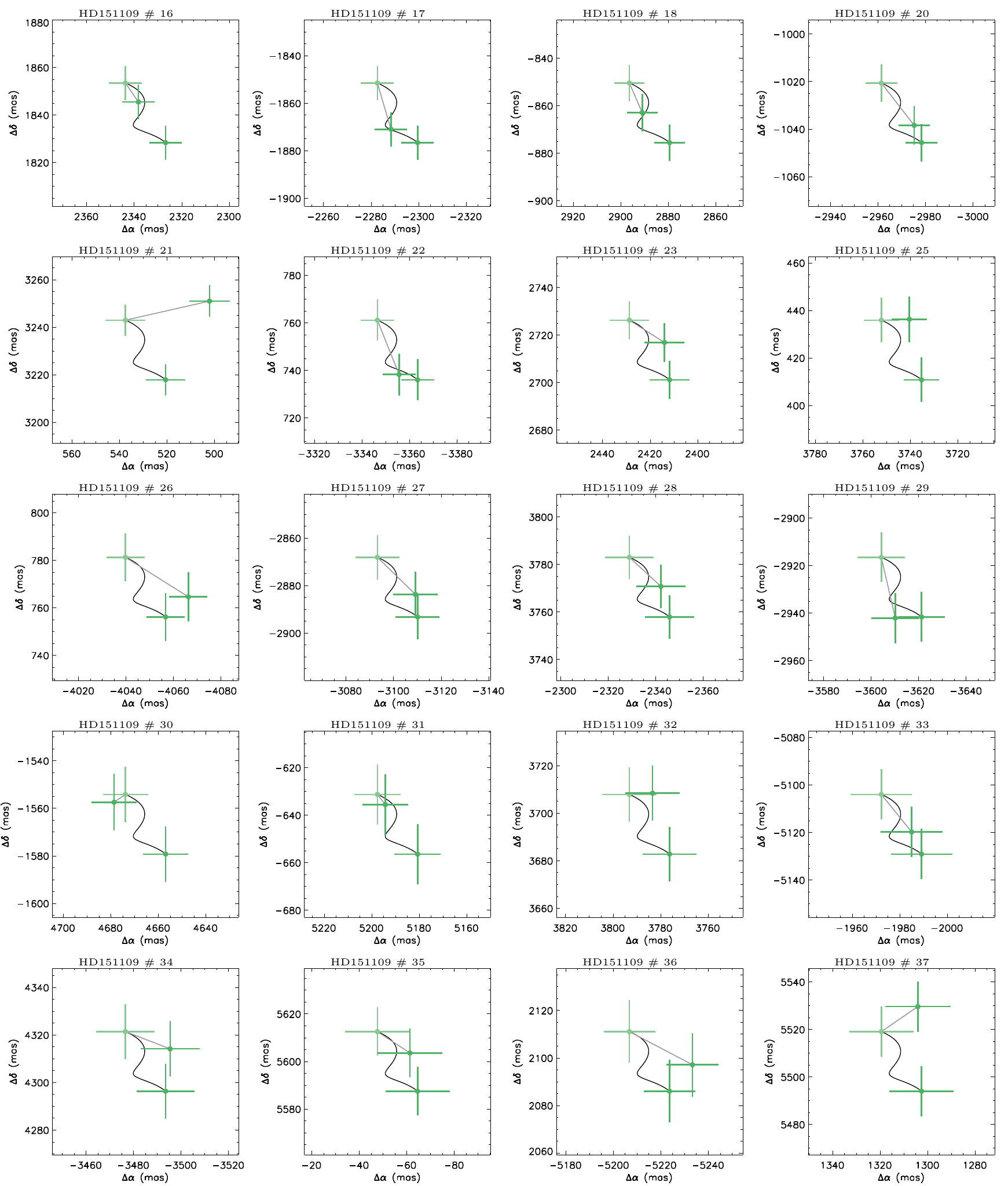}
\end{figure*}

\begin{figure*}
	\includegraphics[width=\textwidth]{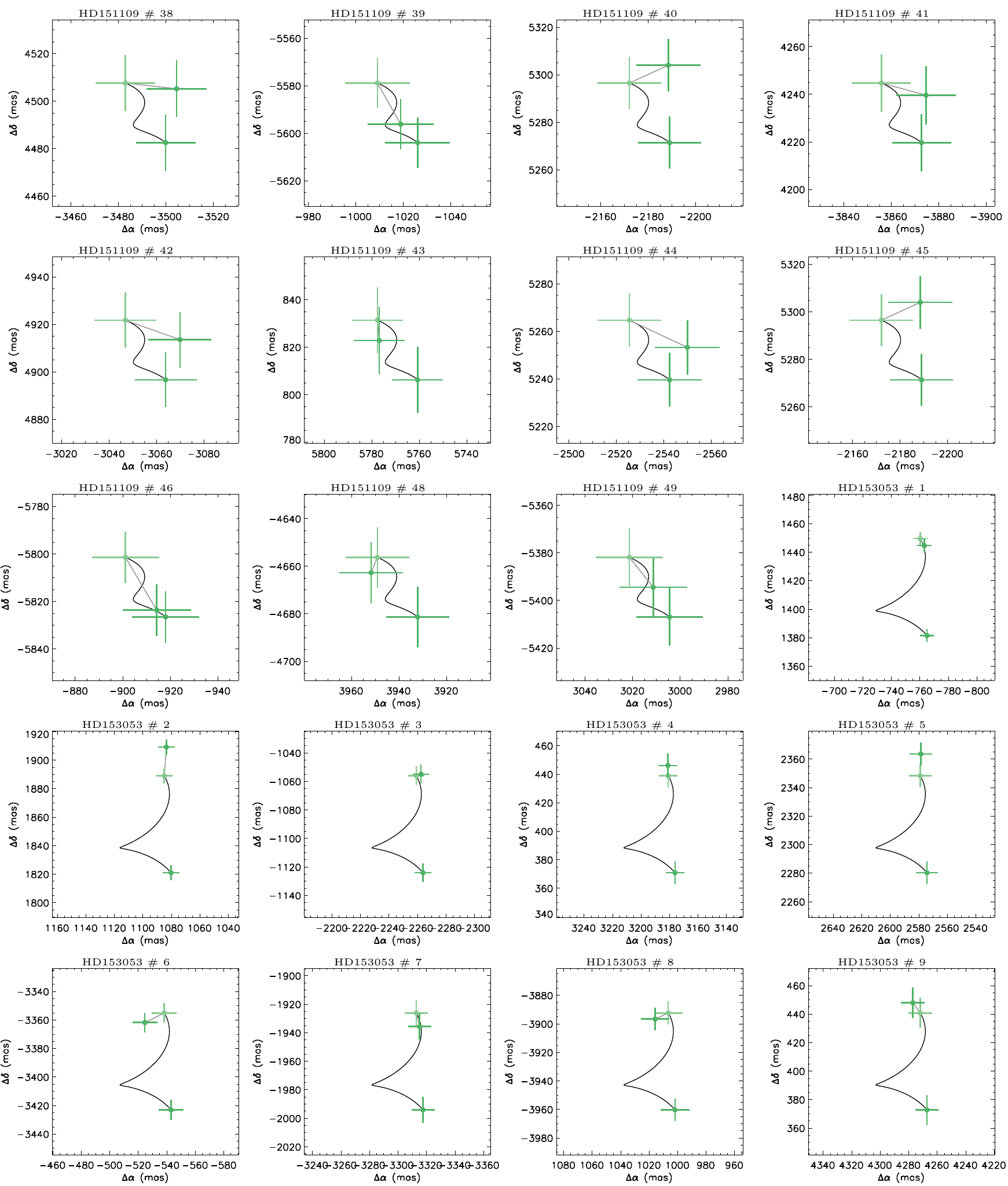}
\end{figure*}

\begin{figure*}
	\includegraphics[width=\textwidth]{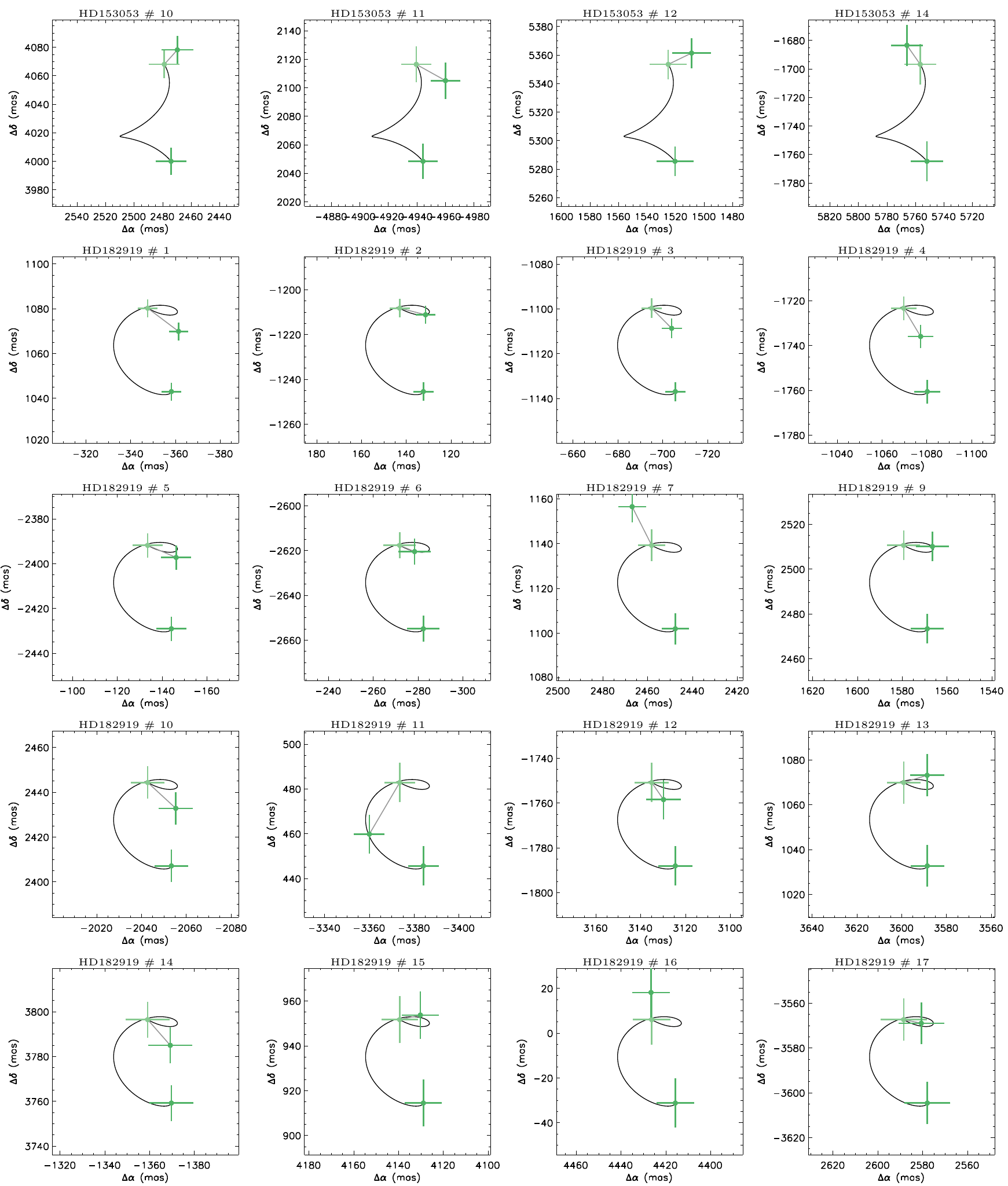}
\end{figure*}

\begin{figure*}
	\includegraphics[width=\textwidth]{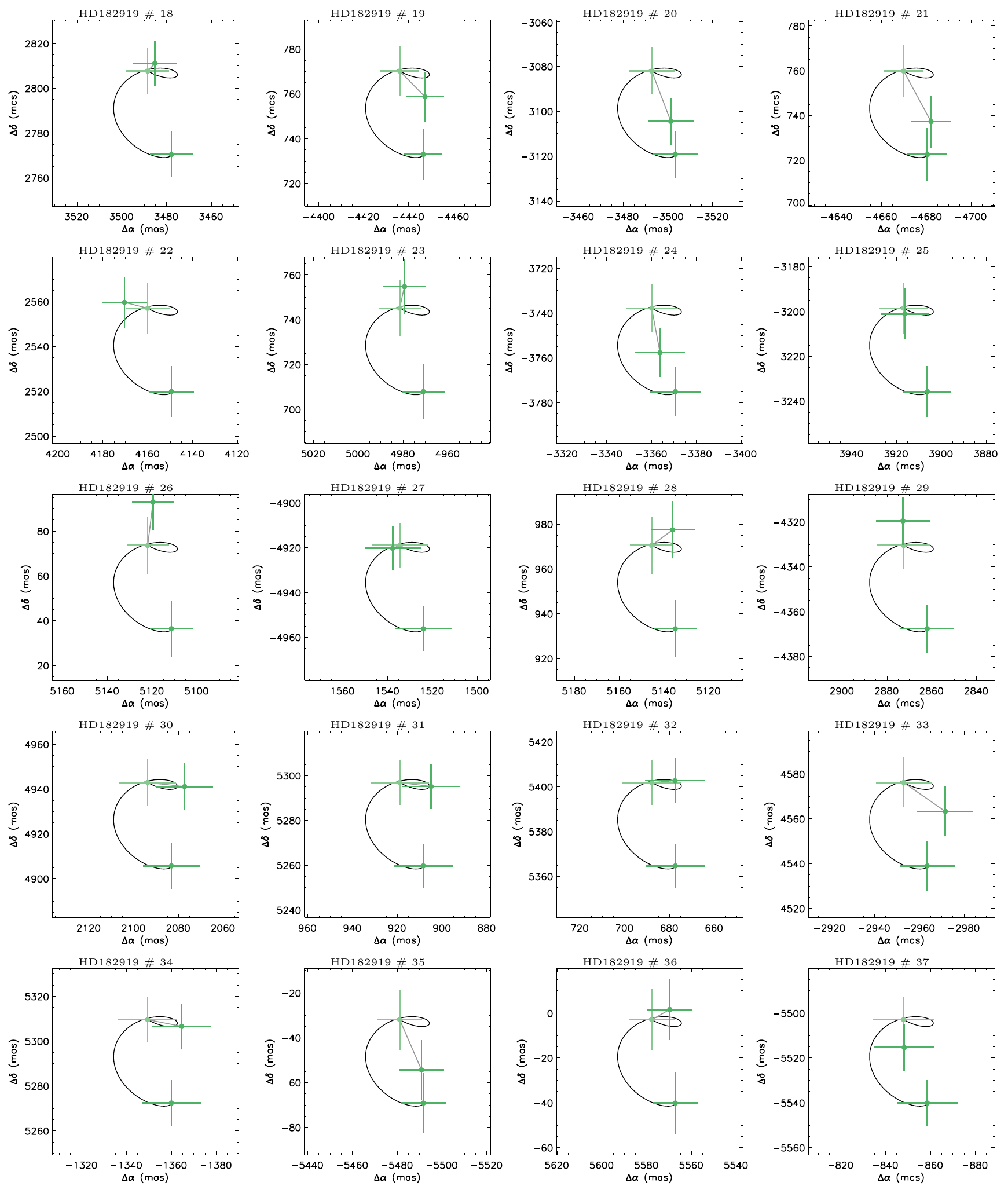}
\end{figure*}

\begin{figure*}
	\includegraphics[width=\textwidth]{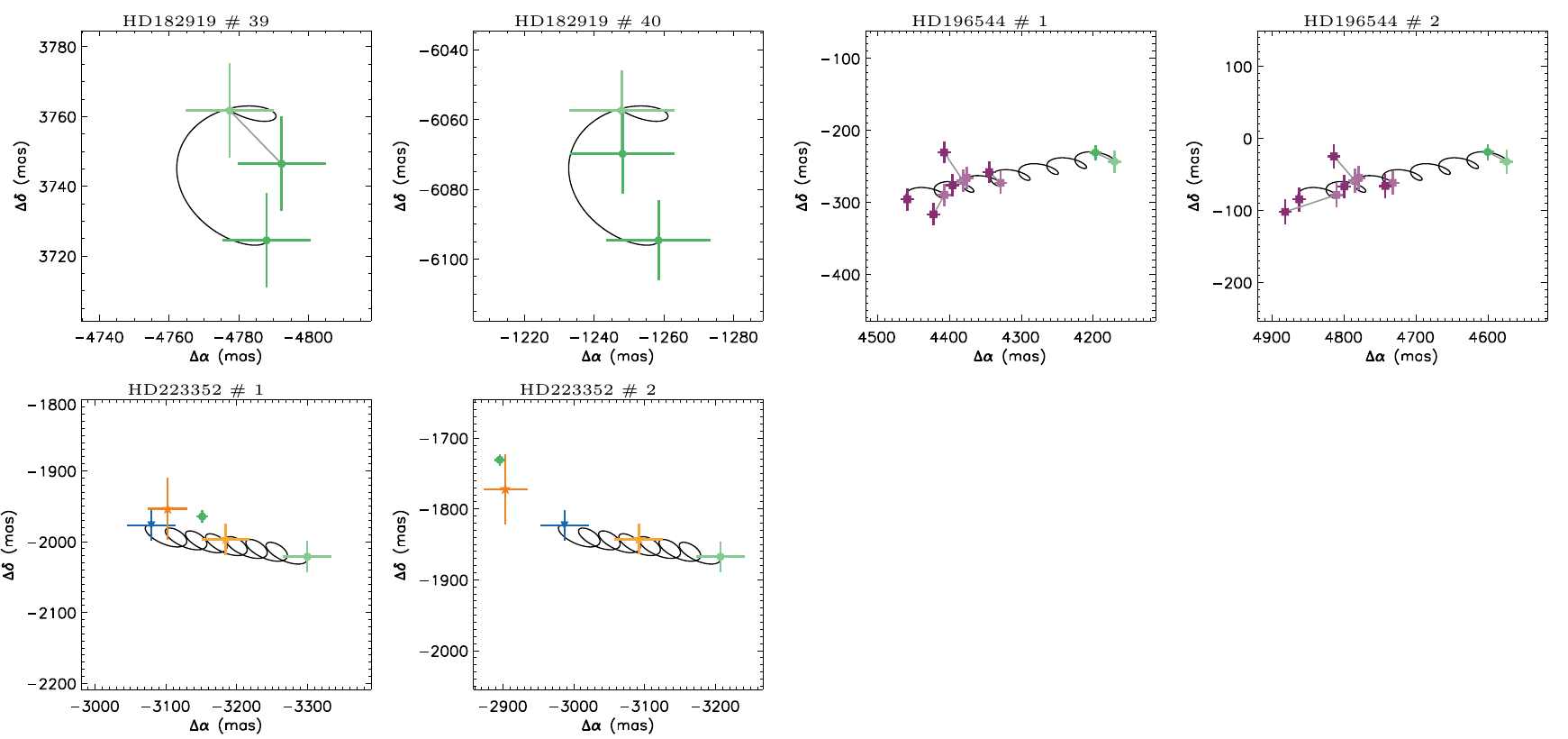}
\end{figure*}

\clearpage

\section{Contrast limits for Individual Observations}
\label{app:contrasts}

Here we present contrast limits for each individual observation. Orange and blue lines represent data from the SPHERE/IRDIS and SPHERE/IFS subsystems respectively. The contrasts are calculated by injecting fake planets, with full details given in Section \ref{sec:contrasts} in the main text.

\begin{figure*}
	\includegraphics[width=0.95\textwidth]{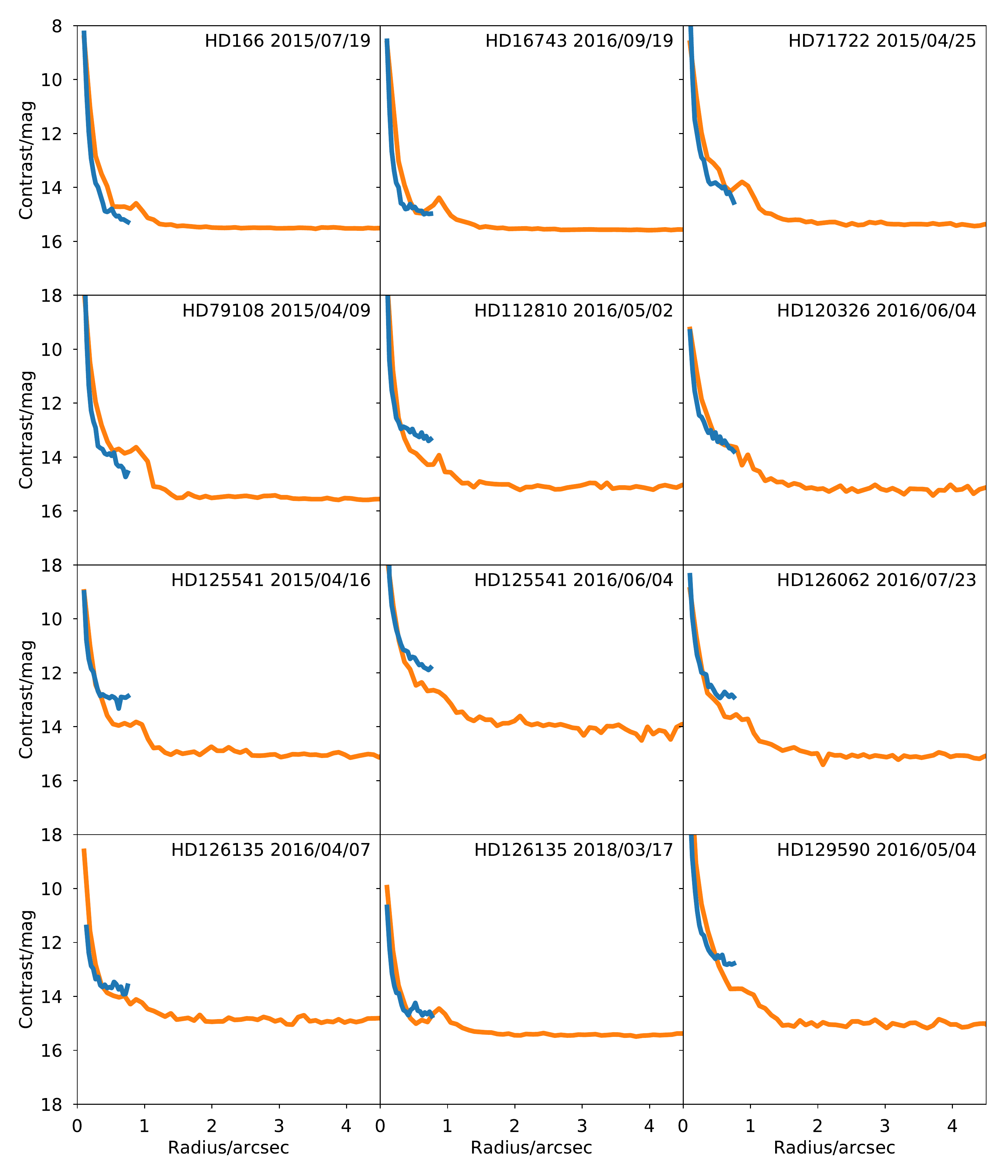}
\end{figure*}

\begin{figure*}
	\includegraphics[width=0.95\textwidth]{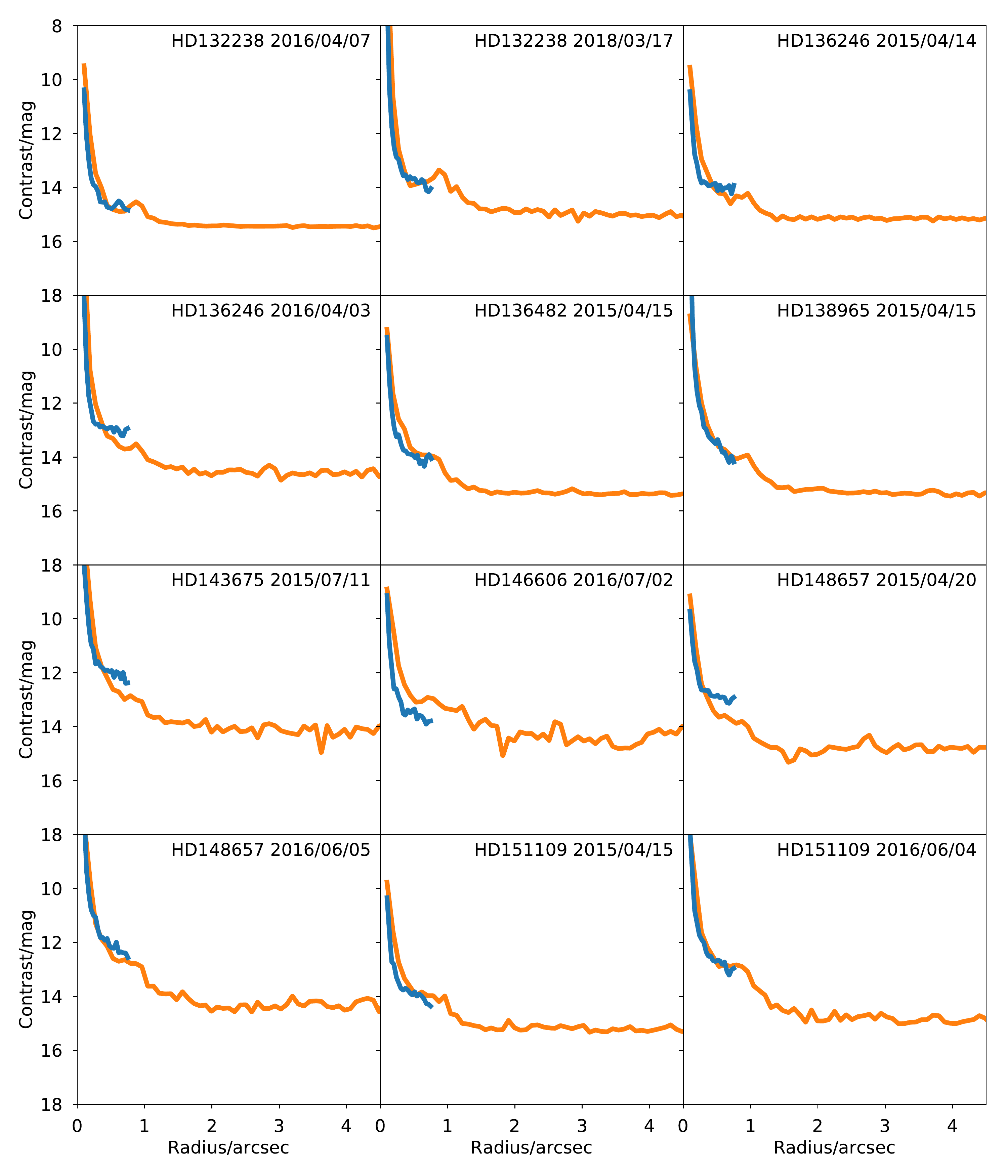}
\end{figure*}

\begin{figure*}
	\includegraphics[width=0.95\textwidth]{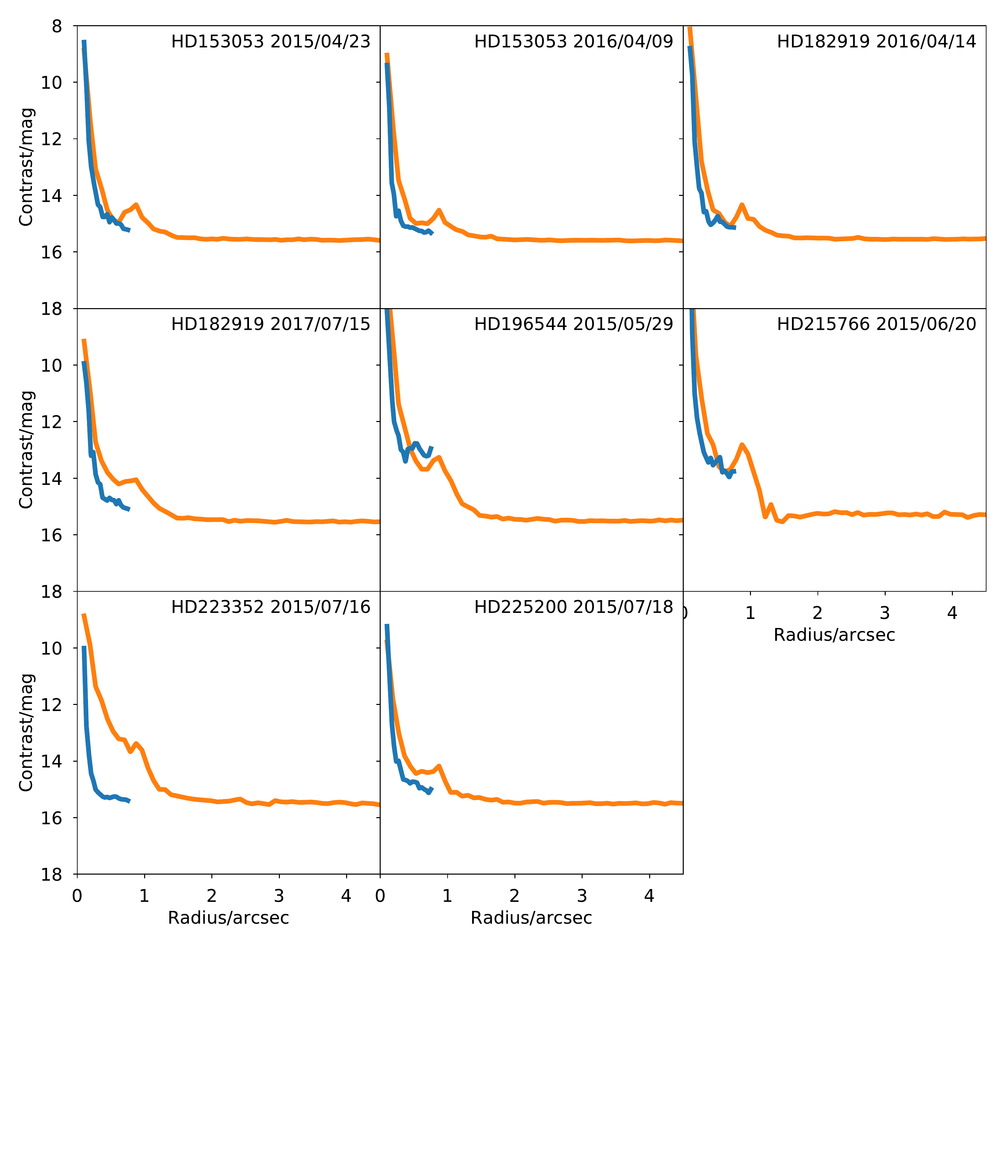}
\end{figure*}

\bsp
\label{lastpage}
\end{document}